\documentclass[10pt,conference,letterpaper]{IEEEtran}
\usepackage{amsmath}
\usepackage{graphicx}
\usepackage{cite}
\usepackage{color}
\usepackage{url}
\newtheorem{proposition}{\bf   Proposition}
\newtheorem{theorem}{\bf Theorem}

\newtheorem{corollary}{\bf Corollary}
\newtheorem{lemma}{\bf Lemma}

\IEEEoverridecommandlockouts

\begin{document}

%\sloppy

\title{
The Impact of Unlicensed Access on Small-Cell Resource Allocation \thanks{This work was
supported by NSF under grant 1343381.}
%A Pricing Model for Optimizing Femto-Cell Deployments With Additional Unlicensed Access
%\thanks{This work was supported by NSF under grant 1343381.}
}

\author{\IEEEauthorblockN{Cheng Chen, Randall A. Berry, Michael L. Honig}
\IEEEauthorblockA{Dept. of EECS, Northwestern University \\
Evanston, IL 60208\\
cchen@u.northwestern.edu, \{rberry, mh\}@eecs.northwestern.edu}
\and
\IEEEauthorblockN{Vijay G. Subramanian}
\IEEEauthorblockA{Dept. of EECS, University of Michigan\\
Ann Arbor, MI 48109\\
vgsubram@umich.edu}
}
\maketitle

\begin{abstract}
%%%Short version
Small cells deployed in licensed spectrum and unlicensed access
via WiFi provide different ways of expanding wireless services
to low mobility users. That reduces the demand for conventional
macro-cellular networks, which are better suited for wide-area
mobile coverage. The mix of these technologies seen in practice
depends in part on the decisions made by wireless service providers
that seek to maximize revenue, and allocations of licensed and
unlicensed spectrum by regulators. To understand these interactions
we present a model in which a service provider allocates available
licensed spectrum across two separate bands, one for macro- and
one for small-cells, in order to serve two types of users: mobile and fixed.
We assume a service model in which the providers can charge
a (different) price per unit rate for each type of service
(macro- or small-cell); unlicensed access is free.
With this setup we study how the addition of unlicensed spectrum
affects prices and the optimal allocation of bandwidth
across macro-/small-cells. We also characterize the optimal fraction
of unlicensed spectrum when new bandwidth becomes available.

\end{abstract}

\section{Introduction}
Current cellular networks are evolving towards
heterogeneous networks (HetNets) to cope with the accelerating demand for wireless data
along with variations in mobility and service requirements\cite{HetNet1-Qualcomm11, HetNet2-Ghosh12}. The primary feature of a HetNet
is the deployment of multiple types of access infrastructure with
different transmission powers, range, and spectral efficiency, such as small-cells targeting
local ``hot spots." In addition to cellular infrastructure in licensed spectrum allocations,
unlicensed services (e.g., WiFi) are  an increasingly common
alternative for providing users local access. As unlicensed networks and HetNets proliferate,
wireless users have greater choice in the type of network
they can access. That can in turn affect strategic decisions
on the part of the SPs regarding pricing and network resource allocation.

%it is expected that HetNets will also make more extensive use of unlicensed spectrum
%for off-loading traffic in licensed bands, and possibly to support stand-alone networks based
%on WiFi or LTE (i.e., LTE-U and LTE-LAA \cite{LTE-U}).
%The interplay of HetNets in licensed spectrum and WiFi networks in unlicensed spectrum
%is not well understood.

While the introduction of HetNets, and small-cell networks in particular, will
increase overall data capacity, network management and resource allocation become
more complicated. In particular, an SP must allocate available resources
across different cell types (small vs macro) taking into account mobility patterns
and demand for different services. That allocation then interacts with pricing strategies
that can differentiate among service categories and controls revenue.
Resource allocation is further complicated by the existence of unlicensed WiFi networks,
which can be viewed
both as an additional resource for offloading traffic, and as a source of competition
for small-cell networks in licensed spectrum.\footnote{For example, AT\&T's WiFi network helps
to expand the capacity of AT\&T's cellular network, whereas Comcast's WiFi service
effectively competes with cellular services.}

In this paper we study the effect of unlicensed spectrum on resource allocation
in HetNets containing large- (macro-) and small- (pico- or femto-) cells.
The macro- and small-cells are assumed to be operated by a cellular SP,
or multiple competing SPs, using licensed (proprietary) spectrum.
To model the demand for different services, we assume two types of users:
mobile and fixed. The mobile users must be served by macro-cells only, whereas
the fixed users can be served by either the macro- or small-cells, or via the unlicensed spectrum.
%In this paper, we consider a heterogeneous wireless network consisting of both licensed and unlicensed spectrum.  In each band of licensed spectrum we assume that service providers build out a two-tier network including macro-cells and small-cells. Mobility of the users is used to model in a simple manner the available user diversity, with fixed users being able to access both small- and marco-cells and the mobile users restricted to the macro-cells.
We assume that the SPs charge a price per unit rate to access their network.
The macro- and small-cells are viewed as providing two different services,
hence the SPs set two corresponding prices for access.
There is no access fee for the unlicensed band.
In all cases the available rate is split among all users sharing the band.
Given this setup, the SPs wish to set prices and allocate bandwidth
across the macro- and small-cells to maximize their revenue.

Our model is similar to that presented in our previous work \cite{Competition5-Chen15} to study
bandwidth allocation in HetNets with competing SPs; however, here
the main distinguishing feature
is the presence of the unlicensed band. Our results show how the
unlicensed spectrum affects the SPs' willingness to allocate resources
to the small-cell network. Moreover, the model and results can be used
to quantify the utility (total welfare) offered by the unlicensed spectrum,
and to compare it with that generated by the licensed spectrum
taking the strategic decisions of the SPs into account.
The introduction of unlicensed spectrum raises several analytical challenges
\emph{vis-a-vis} prior work (e.g., \cite{Competition5-Chen15}).
This is due to the expanded set of choices (i.e., for fixed-user association)
and also the additional conditions on equilibria due to competition from the
unlicensed service.

We now summarize our main results. We first focus on a monopoly SP, and then
consider the more complicated scenario with multiple competing SPs.
In all scenarios we model the SP and user actions as a two-stage game in which
the SPs first partition the licensed band into subbands for
the macro- and small-cell networks, and subsequently announce prices for services.
The prices then enable the fixed users to determine their network association
(macro-/small-/unlicensed). The following results are obtained by analyzing
sub-game perfect equilibria, assuming a class of $\alpha$-fair utility functions
for tractability.

1. \emph{HetNet Market structure:} In equilibrium the macro-cell network serves {\em only}
mobile users. Fixed users then associate with either the small-cell or unlicensed network.
This applies when the SPs maximize either revenue or social welfare. Hence this association
is optimal in the sense of maximizing social welfare.\footnote{This result was also shown
to hold without unlicensed spectrum in \cite{Competition5-Chen15}.}

2. \emph{Bandwidth allocation with monopoly SP:}
Comparing bandwidth allocations with and without unlicensed spectrum,
when maximizing revenue, a monopoly SP may allocate {\em more} bandwidth
to small-cells with unlicensed than without.
This occurs when the unlicensed network offers a sufficiently low rate
(e.g., due to a small amount of unlicensed bandwidth).
This seems counterintuitive when
the unlicensed network is viewed as an additional resource; however,
this is because the presence of unlicensed bandwidth alters the dependence of
quantity (users served) with price.
In contrast, when the SP maximizes social welfare, it always
allocates less bandwidth to small cells with unlicensed spectrum.

%With a monopoly service provider,  we characterizes the unique bandwidth allocation
%for both revenue and social welfare maximization. Intuitively, considering competition
%from unlicensed access, it appears that compared with the scenario without
%unlicensed spectrum, the service provider should invest less bandwidth to small-cells
%in order to maximize its revenue. However, we show that actually this is not necessarily
%the case. For revenue maximization, in contrary, when the total rate in unlicensed spectrum
%is small, the service provider needs to invest more bandwidth small-cells compared
%with the case without unlicensed spectrum. It is only when the total rate in
%unlicensed spectrum is sufficiently large that the service provider would decrease
%the bandwidth in small-cells. Nonetheless, for social welfare maximization
%the service provider should always invest less bandwidth to small-cells relatively.

3. \emph{Equilibrium with competitive SPs:}
With multiple competitive SPs we prove that there always exists a unique
sub-game perfect equilibrium with an associated bandwidth allocation.
Furthermore, the equilibrium can fall in one of three categories:
(1) all SPs allocate bandwidth to both macro- and small-cells
(``Macro-Small Nash Equilibrium'', or MSNE);
(2) a subset of SPs allocate bandwidth to macro-cells only,
and the rest allocate to both macro and small cells
(``Macro-Preferred Nash Equilibrium'', or MPNE); and
(3) all SPs allocate bandwidth only to macro-cells
(``Macro-only Nash Equilibrium'', or MNE).
Without unlicensed spectrum for our choice of utility functions only the MSNE is possible \cite{Competition5-Chen15}.
Hence the increased competition from the unlicensed network may cause
some SPs to give up on small-cells and allocate bandwidth only to the macro-cells.
We also consider the asymptotic scenario where the number of SPs goes to infinity
and observe that in general, the equilibrium with an arbitrary amount
of unlicensed spectrum does not achieve the maximum social welfare.

4. \emph{Dependence of social welfare on the licensed/unlicensed split:}
For spectrum regulators, such as the FCC, a challenge is determining
how much newly available spectrum should be licensed or unlicensed.
We use our results to illustrate how the mix of licensed
and unlicensed spectrum affects social welfare.
This depends crucially on the relative spectral efficiencies associated
with the small-cell and unlicensed networks. If the small-cell network
has higher spectral efficiency, then allocating the entire bandwidth as licensed
maximizes social welfare. Conversely, if the unlicensed network has
higher spectral efficiency, then there is an optimal (positive) amount
of unlicensed spectrum, which may or may not achieve
the maximum social welfare.
Furthermore, allocating insufficient unlicensed bandwidth in this scenario
can cause the social welfare to decrease below the all-licensed allocation.
(This is illustrated in Fig. 5 in Sec. VII.)

%With additional unlicensed spectrum, we find that for both revenue and social welfare maximization, the optimal market structure is still one where macro-cells only serve mobile users and femto-cells only serve (a subset of) fixed users, with the remaining fixed users choosing unlicensed access. With a monopoly service provider, for revenue maximization the amount of bandwidth allocation to femtocells depends on the rate capacity of WiFi network. When the rate capacity in unlicensed spectrum is small, the service provider should allocate less bandwidth to femtocells, while when it is large the service provider should increase the bandwidth in femtocells. In contrast, for social welfare maximization, the service provider should always allocate less bandwidth to femtocells, no matter the rate in the unlicensed spectrum. With multiple competitive service providers, a unique Nash equilibrium exists, and is three different types: {\color{red} Femto-Macro Nash Equilibrium (FMNE) where ...; Macro-Preferred Nash Equilibrium (MPNE) where ...; and Macro-only Nash Equilibrium (MNE) where ....} Moreover, we also characterize the properties of different types of Nash equilibria. For example, at an FMNE, compared with the case without unlicensed spectrum, SPs would always invest less total bandwidth to femto-cells, independent of the total rate available in unlicensed spectrum.

Related work on
%There have been two different models studying
pricing and bandwidth allocation in heterogeneous wireless networks
is presented in \cite{Opt1-Shetty09,Opt2-Gussen11,Opt3-Yun12,Opt4-Chen11, Opt5-Lin11, Opt6-Duan13, Opt7-Chen13}.
In \cite{Opt1-Shetty09,Opt2-Gussen11,Opt3-Yun12}
small-cell service is an enhancement to macro-cell service,
and in \cite{Opt4-Chen11, Opt5-Lin11, Opt6-Duan13, Opt7-Chen13}
small- and macro-cells provide separate services (as we assume here).
Optimal pricing only is studied in \cite{Opt1-Shetty09, Opt3-Yun12},
whereas joint pricing and bandwidth allocation is studied
in \cite{Opt2-Gussen11, Opt4-Chen11, Opt5-Lin11, Opt6-Duan13, Opt7-Chen13}.
However, in that work there is a single SP (monopoly) and no unlicensed spectrum.
%All of these analyze the case of a single service provider, i.e., the monopolistic setting, and do not model any interactions with unlicensed access.

In \cite{Competition1-Zhang13, Competition2-Hossain08, Competition3-Sengupta07,
Competition4-Zhang08, Competition5-Chen15}, competition among multiple SPs providing
HetNet services is investigated. References
\cite{Competition1-Zhang13, Competition2-Hossain08} study pricing and service competition
with fixed bandwidth allocations, while
\cite{Competition3-Sengupta07, Competition4-Zhang08, Competition5-Chen15}
study both pricing and bandwidth optimization.
However, in \cite{Competition3-Sengupta07, Competition4-Zhang08}
the SPs compete to acquire the spectrum from a spectrum broker,
as opposed to optimizing the bandwidth allocation across the
different cell types in \cite{Competition5-Chen15}.
The preceding work does not consider any interactions with unlicensed access.

While the preceding work focuses on HetNet deployments using licensed spectrum,
the interaction of unlicensed with licensed spectrum in other contexts is considered in
%but without a HetNet architecture
\cite{LicensedUnlicensed1-Bazelon09, LicensedUnlicensed2-Maille09, LicensedUnlicensed3-Nguyen14}.
In \cite{LicensedUnlicensed1-Bazelon09} an economic analysis of the trade-off
between incremental licensed and unlicensed spectrum allocations is presented,
which shows that licensed spectrum is the favored option.
In \cite{LicensedUnlicensed2-Maille09}, an intermediate model of price competition
between two SPs having a fixed licensed part of the spectrum
and sharing the remaining part as unlicensed is proposed.
It is shown that user welfare increases with the proportion of unlicensed spectrum;
however, the overall social welfare decreases, indicating resources are used
less efficiently.
%also investigate the influence of the shared unlicensed band
%on social and user welfare and show that when the proportion of
%unlicensed spectrum increases, the user welfare increases. However,
Reference \cite{LicensedUnlicensed3-Nguyen14} studies social welfare when
unlicensed spectrum is added to an existing allocation of licensed spectrum
among incumbent competing SPs. The conclusion is that the social welfare can
%depends on the amount of additional unlicensed spectrum; in fact, social welfare
decrease over a significant range of unlicensed bandwidths.
%When users have different sensitivities to congestion (delay), it is
%shown that user surplus (in addition to total welfare) can also decrease.
%It also considers heterogeneous users and characterizes user surplus as well as social welfare.
%The results shows that with heterogeneous users the adding unlicensed spectrum
%can also decrease user surplus.
The preceding work assumes that the licensed spectrum supports
a single type of service, hence does not consider the associated
bandwidth allocation problem studied here. Also, except for
\cite{LicensedUnlicensed3-Nguyen14}, the users are assumed to be homogeneous.

%%%Previous submission paragragh%%%
The paper is organized as follows. We introduce our system model in
Section \ref{Sec:System Model}. The price and user association equilibrium for a
monopoly SP is presented in Section \ref{Sec:Price and User Association Equilibrium}.
Optimal bandwidth allocation for revenue maximization is discussed in
Section \ref{Sec:Optimal Bandwidth Allocation}, and for social welfare maximization
in Section \ref{Sec:Social Welfare Maximization}. We then consider multiple
competing SPs in Section \ref{Sec:Service Competition Among Multiple Service Providers}.
The dependence of social welfare on the amount of unlicensed bandwidth
is discussed in \ref{Sec:Licensed vs Unlicensed Spectrum}, and
%Numerical results are presented in Section \ref{Sec:Numerical Results} and
conclusions are presented in Section \ref{Sec:Conclusion}. All
proofs of the main results and several supplemental results can be be found in the appendices.

%%%Final verstion paragragh, shortened for space limitation%%%
%The paper is organized as follows. We introduce our system model in
%Section \ref{Sec:System Model}. Monopoly scenario with a single revenue-maximizing
%SP is discussed in Section \ref{Sec:Price and User Association Equilibrium}
%and \ref{Sec:Optimal Bandwidth Allocation}, whereas for social welfare maximization in
%Section \ref{Sec:Social Welfare Maximization}. We then consider multiple
%competing SPs in Section \ref{Sec:Service Competition Among Multiple Service Providers}.
%The dependence of social welfare on the amount of unlicensed bandwidth
%is discussed in \ref{Sec:Licensed vs Unlicensed Spectrum}.
%We conclude in Section \ref{Sec:Conclusion}. Due to space considerations, all
%proofs can be be found in an on-line full version of this paper \cite{FullVersionWithAppen}.

\section{System Model}\label{Sec:System Model}
\subsection{Network Model}

We consider a scenario where there are one or more SPs, each of which has a two-tier cellular network operating in licensed spectrum consisting of macro-cells (with wide coverage) and small-cells (with only local coverage).  Macro-cells and small-cells are assumed to operate in separate
bands.\footnote{Equivalently, macro- and small-cells could operate in different time-slots, e.g., using the Almost Blank Subframes (ABS) feature in LTE \cite{ABS}.}  We also assume that there is unlicensed spectrum in which WiFi Access Points (APs) are deployed with no access charges. The assumption of no access charge is made to make the unlicensed as desirable as possible to the users. Users in the network are classified into two categories based on their mobility: mobile users that can only be served by macro-cells and fixed users that can be served by macro-cells, small-cells or WiFi (but not multiple at the same time).

Macro-cells, small-cells and WiFi APs are assumed to be uniformly deployed over a given area.\footnote{Alternatively, we can view small-cells and WiFi APs as being uniformly
deployed over ``hot spot" areas and restrict fixed users to these areas.}
To simplify our analysis we restrict our attention to the case where all SPs have the same infrastructure deployment density. The density of macro-cells per SP is normalized to one and we denote the densities of small-cells and WiFi APs by $N_S$ and $N_U$, respectively. Both types of users are also uniformly deployed over the given area, and we assume a large number of users so that we can model them as non-atomic. The density of fixed and mobile users are given by $N_f$ and $N_m$, respectively.

We assume that each SP $i$ has total bandwidth $B_i$ and the macro-cells of SP $i$ can provide a total data rate of $C_{i, M}=B_{i, M}R_0$, where $R_0$ is the (average) spectral efficiency and $B_{i,M}$ is the bandwidth allocated to macro-cells by SP $i$.\footnote{Of course the actual rate a SP can provide at any time will depend on many factors such as the channel gains to its users and the scheduling algorithm employed. Here, we view $C_i$ as averaging over such effects over a long enough time horizon, which is reasonable for the network planning problems we consider.}
Similarly, the total available rate in small-cells of SP $i$ and in WiFi APs are defined as $C_{i,S}=\lambda_S B_{i, S}R_0$ and $C_U=\lambda_U B_UR_0$, respectively. Here $\lambda_S$ and $\lambda_U$ reflect the rate difference due to the combination of spectral efficiency and deployment density differences in small-cells and WiFi APs compared with macro-cells. Since small-cells generally have a higher spectral efficiency and larger deployment density, we assume $\lambda_S>1$. WiFi APs typically have a higher deployment density but lower spectral efficiency. Therefore, we don't make any assumptions on $\lambda_U$.

\subsection{Market Model}

Each SP $i$ offers separate macro- and small- cell service. It charges all users of a given service the same price per unit rate, which are denoted by $p_{i, M}$ and $p_{i, S}$ for macro- and small- cell service of provider $i$, respectively. Of course, other pricing approaches arise in practice such as flat-rate pricing. We focus on per unit pricing in part because it is analytically tractable and in part because, as will be shown in our analysis, it is sufficient for optimizing the total user welfare. Recall, WiFi service is assumed to have no access charge.

The density of users connected to macro-cells and small-cells of SP $i$ and to WiFi are denoted by $K_{i, M}$, $K_{i, S}$ and $K_U$, respectively. Note that $K_{i, S}$ and $K_U$ only consist of fixed users, while $K_{i, M}$ may include both mobile users and fixed users. Additionally, mobile users are assumed to have priority connecting to macro-cells. As a result, macro-cells can only accommodate fixed users if all mobile users have been served.

\subsection{User Optimization}

Each user is endowed with a utility function $u(r)$, which only depends on the service rate it gets from any type of service. For simplicity of analysis, we assume that all users have the same utility and we further restrict this to be an $\alpha$-fair utility functions \cite{MoWalrand} with $\alpha \in (0,1)$:
\begin{equation}\label{Eqn:UtilityFn}
u(r)=\frac{r^{1-\alpha}}{1-\alpha}, \quad \alpha \in (0,1).
\end{equation}
\noindent This restriction enables us to explicitly calculate many equilibrium quantities, which appears to be difficult for more general classes of utility. Further this class is widely used in both networking and economics, where it is a subset of the class of iso-elastic utility functions.\footnote{In general $\alpha$-fair utilities require that $\alpha \geq 0$ to ensure concavity; requiring $\alpha>0$ ensures strict concavity but allows us to approach the linear case as $\alpha \rightarrow 0$.  The restriction of $\alpha<1$  ensures that utility is non-negative so that a user can always ``opt out" and receive zero utility. Note also that as $\alpha \rightarrow 1$, we approach the  $\log(\cdot)$ (proportional fair) utility function.}

For a given licensed service, each user requests a rate that maximizes their net payoff $W$, defined as the difference between their utility and the cost of service, i.e. they solve:
\begin{align}\label{Opt:User Optimization}
%\notag
\underset{r\geq 0}{\text{maximize}} \quad & W = u(r)-pr,
%\\ \text{subject to} \quad & r \ge 0
\end{align}
where as described below, the SPs will set the prices $p$ to ensure that the resulting demand can be
met.\footnote{We assume that the users are {\it price taking} in that they do not anticipate how their selection of service or rate will effect the resulting prices, which is reasonable under our assumption of
many small users.} For WiFi service, since there is no access charge, we assume all users simply share the available total rate equally.
Since fixed users can choose any one service, they would choose the one with the largest payoff.

For $\alpha$-fair utility functions,  (\ref{Opt:User Optimization}) has the  unique solution:
\begin{equation}\label{Eqn:UserRateOpt}
r^{*}= D(p)=(u^{\prime})^{-1}(p)=(1/p)^{1/\alpha},%\frac{1}{\sqrt[\alpha]{p}}%\max \Big((u^\prime)^{-1}(p),0 \Big)
\end{equation}
where $D(p)$ is the user's rate demand function and $u^{\prime}$ denotes the first derivative of $u$. The maximum net payoff for a user is thus:
\begin{equation} \label{Eqn:User Net Payoff}
W^*(p) = u(D(p))- pD(p)=\frac{\alpha}{1-\alpha}p^{1-\frac{1}{\alpha}}.
\end{equation}

%For (fixed) users served by WiFi network, since there is no price in unlicensed spectrum, it is therefore assumed all users simply share the available total rate equally.

\subsection{Service Provider Optimization}
Each SP $i$ needs to decide on the bandwidth partition $(B_{i, M}, B_{i, S})$ and pricing decisions $(p_{i, M}, p_{i, S})$ to maximize its revenue $S_i$, i.e., the aggregate amount paid by all users choosing its macro- and small- cell services. This can be formulated as:
\begin{align}\label{Opt:SP Optimization}
\notag \text{maximize} \quad & S_i =  K_{i, M}p_{i, M}D(p_{i, M})+K_{i, S}p_{i, S}D(p_{i, S})\\
\notag \text{subject to} \quad & K_{i,M}D(p_{i, M}) \leq C_{i,M}, K_{i,S}D(p_{i, S}) \leq C_{i,S}\\
\notag &  0\le p_{i, M}, p_{i, S}<\infty  \\
&B_{i, M}, B_{i, S}\ge 0, B_{i, M}+B_{i, S} \le B_i.
\end{align}
Here the first constraint ensures that the SP can meet the rate demanded by the users, where recall $C_{i,M}$ and $C_{i,S}$ depend on the bandwidth allocation. Note also that $K_{i,M}$ and $K_{i,S}$ will depend on the users associations, which in the case of multiple SPs will depend on the prices and bandwidths chosen by those SPs. Hence, we model the choice of bandwidth allocations and prices as a game played among the SPs. In practice, since bandwidth allocation takes place over a slower time-scale than price adjustments, we view this game as consisting of the  following two stages: First, SPs determine their bandwidth allocation between macro-cells and small-cells. Then given their bandwidth allocation, SPs announce prices for both macro- and small-cells. Users then choose the services accordingly based on our association rules.

For such a game we characterize its sub-game perfect equilibrium by first characterizing a {\it user association equilibrium} in which SPs set prices given a fixed bandwidth allocation and then study
the equilibrium bandwidth allocation based on the results obtained in the first step.

We also consider the choice of bandwidth allocations and prices that maximize the social welfare, i.e. the sum utility over all users. This can be formulated as maximizing
\begin{equation}\label {Opt:SW optimization}
\text{SW}= \sum\limits_{i=1}^{N}K_{i, M}u(R_{i, M})+K_{i, S}u(R_{i, S})+
K_U u(R_U)
\end{equation}
subject to the same constraints for each SP $i$ as in (\ref{Opt:SP Optimization}).
%That problem is stated as:
%\begin{align} \label {Opt:SW optimization}
%\notag \text{maximize}  \quad &\text{SW}= \sum\limits_{i=1}^{N}K_{i, M}u(R_{i, M})+K_{i, S}u(R_{i, S})+\\
%&K_U u(R_U)\\
%\notag \text{subject to}\quad &   0\le p_{i, M}, p_{i, S}<\infty\\
%&B_{i, M}, B_{i, S}\ge 0, B_{i, M}+B_{i, S}\le B_i
%\end{align}
Here, $R_{i, M}$, $R_{i, S}$ and $R_U$ denote the average service rates per user in each respective service, which in turn depends on the prices and bandwidth allocations.

\section{User Association with a Single SP}\label{Sec:Price and User Association Equilibrium}
In this section we study the user association equilibrium
given a fixed bandwidth allocation between macro- and small-cells by a single SP.
Here we assume the SP maximizes revenue, and consider social welfare maximization
in Section \ref{Sec:Social Welfare Maximization}. Since mobile users can only connect to macro-cells,
we need only consider the association of fixed users.
%In our model fixed users have three choices: macro-cells, small-cells in licensed spectrum and WiFi network in unlicensed spectrum. If fixed users associate with macro-cells or small-cells, they need to pay the cost based on the price per unit rate set by the service provider. However, if fixed users connect to WiFi network, there is no usage fee.

Given a fixed bandwidth allocation and annoucned prices, users select their service to maximize their
net payoff.
For macro- and small-cell service the net payoff
for a user is given by (\ref{Eqn:User Net Payoff}).
Since
%For $\alpha$-fair utility functions,
$W^*(p)$ is a decreasing function of $p$, the choice between macro- and small-cell networks,
ignoring the unlicensed network,
is simply the one with the lower price.
For the WiFi service, the users net payoff is simply given by $u(\frac{C_U}{K_U})$.
%fixed users would always prefer the one with lower price. More thought is needed for the choice between macro-cells (or small-cells) and WiFi network: fixed users need to compare the net payoff and choose the one with larger payoff.
With non-atomic users, it follows that in any user association equilibrium the net payoffs of any
services that are used by fixed users must be equal, and any service not used must have a lower
net payoff.

The SP adjusts the prices $p_M$ and $p_S$ to maximize its revenue
taking into account the user association equilibrium. \footnote{Here we drop the SP subscript $i$.} 
Two scenarios are possible:\\
1. {\em Mixed service}: Macro-cells serve both mobile users and a subset of fixed users; \\
2. {\em Separate service}: Macro-cells only serve mobile users.
%The next theorem states the condition for each type of service.

%1. Mixed service scenario: Macro-cells serve all mobile users and a subset of fixed users jointly. Small-cells and WiFi network serve the remaining fixed users. Prices in macro-cells and small-cells are the same and this price equalizes the total rate demand and total rate supply in licensed spectrum.
%2. Separate service scenario: Macro-cells only serve mobile users. All fixed users are served by small-cells and WiFi network. Macro-cell price is higher than small-cell price and the prices equalize the total rate demand and rate supply in corresponding cells.

%The next theorem characterizes the optimal price choice and user association equilibrium with a single SP. It indicates that when SP invests a lot of bandwidth to macro-cells, some fixed users would finally associate with macro-cells. Otherwise macro-cells only serve mobile users.
\begin{theorem}[User Association]\label{Thm:Price and User Association Equilibrium}
Given a fixed bandwidth allocation between macro- and small-cells,
at the SPs' optimal prices, the market clears, i.e., all users are served, and all rate is allocated.
Further, there exists a threshold $B_{S,0}$ such that if
$
B_S < B_{S,0}
$,
then the mixed service scenario holds. Otherwise, the separate service scenario holds.

For the mixed service scenario, $p_M = p_S$, whereas
for the separate service scenario $p_M > p_S$.
\end{theorem}
\vspace{6pt}
%The next theorem characterizes the optimal price choice and user association equilibrium with a single SP. It indicates that

Hence, fixed users choose to associate with macro-cells only when the small-cell bandwidth
is sufficiently small. In equilibrium
%The specific user association density for the two scenarios can be computed by using the fixed users' association criterion. That is,
the mixed service scenario implies that the net payoff to fixed users from
connecting to macro-, small-cells and WiFi must be the same,
and so we can write
\begin{align}
\begin{split}
u\Big(\frac{C_S}{K_S}\Big)-p_S\frac{C_S}{K_S}& =u\Big(\frac{C_U}{K_U}\Big) =u\Big(\frac{C_M}{K_M}\Big) - p_M\frac{C_M}{K_M},
\label{eq:mixed}
\end{split}
\end{align}
Note here we are using the fact from Theorem~\ref{Thm:Price and User Association Equilibrium} that all of the rate is allocated. For separate service the same equality will hold except only for small-cells and unlicensed. With $\alpha$-fair utilities, the rate per user $\frac{C_S}{K_S}$ must satisfy (\ref{Eqn:UserRateOpt}). Using this in the above equations yields the following lemma that gives a useful comparison of the equilibrium rates of each service.

\begin{lemma}\label{Lemma:Fixed user equilbrium}
%According to fixed users' association criterion,
In equilibrium fixed users in licensed spectrum (macro-/small-cells)
achieve $\frac{1}{\kappa}$ times the average rate of users
in unlicensed spectrum where
$
%\label{eq:kappa}
\kappa=\alpha^{1/(1-\alpha)}
$.
\end{lemma}

Note that $1/\kappa>1$, which accounts for the access price charged in licensed spectrum. Interestingly with $\alpha$-fair utilities this ratio is independent of any system parameters.

Using this quantity $\kappa$ we can express the bandwidth threshold in Theorem~\ref{Thm:Price and User Association Equilibrium} as
\begin{equation}
\label{eq:BS0}
B_{S,0} = \max\left( {\kappa N_fB_MR_0-N_mC_U,0} \right) / ({\kappa N_m\lambda_SR_0} ).
\end{equation}

From Theorem~\ref{Thm:Price and User Association Equilibrium}, since all users are served we have $K_U+K_S+K_M = N_m+N_f$. Using this and Lemma~\ref{Lemma:Fixed user equilbrium}, we have the following explicit expressions for the equilibrium densities of users for each type of service in the mixed service scenario:
\begin{align}\label{Eqn:Mixed Service}
\begin{split}
K_U &=(N_f+N_m)\frac{C_U}{C_U+\kappa(C_M+C_S)},\\
K_M &=(N_f+N_m)\frac{\kappa C_M}{C_U+\kappa(C_M+C_S)},\\
K_S &=(N_f+N_m)\frac{\kappa C_S}{C_U+\kappa(C_M+C_S)}.
\end{split}
\end{align}
%which gives the number of active users in terms of the network capacities.

Similarly, for the separate service scenario, we have:
%\begin{align}
%u\Big(\frac{C_U}{K_U}\Big)&=u\Big(\frac{C_S}{K_S}\Big)-p_S\frac{C_S}{K_S},\\
%K_M&=N_m, K_S+K_U=N_f,\\
%D(p_M)&=R_M=\frac{C_M}{K_M}, D(p_S)=R_S=\frac{C_S}{K_S}.
%\end{align}
%1. Mixed service scenario:
%\begin{align}
%\notag u\Big(\frac{C_U}{K_U}\Big)& =u\Big(\frac{\lambda_SB_SR_0}{K_S}\Big)-p_S\frac{\lambda_SB_SR_0}{K_S}\\
%&=u\Big(\frac{B_MR_0}{K_M}\Big) - p_M\frac{B_MR_0}{K_M}\\
%K_U+K_S+K_M&=N_m+N_f\\
%D(p_M)=D(p_S)&=\frac{(B_M+\lambda_SB_S)R_0}{K_M+K_S}
%\end{align}
%\noindent 2. Separate service scenario:
%\begin{align}
%u\Big(\frac{C_U}{K_U}\Big)&=u\Big(\frac{\lambda_SB_SR_0}{K_S}\Big)-p_S\frac{\lambda_SB_SR_0}{K_S}\\
%K_M&=N_m, K_S+K_U=N_f\\
%D(p_M)&=\frac{B_MR_0}{K_M}, D(p_S)=\frac{\lambda_SB_SR_0}{K_S}
%\end{align}
%\noindent 2. Separate service scenario:
\begin{align}\label{Eqn:Separate Service}
\begin{split}
K_M&=N_m, \quad K_S=N_f\frac{\kappa C_S}{\kappa C_S+C_U},\\
K_U&=N_f\frac{C_U}{\kappa C_S+C_U}.
\end{split}
\end{align}

\section{Bandwidth Allocation with a Single SP}\label{Sec:Optimal Bandwidth Allocation}
We now consider optimal bandwidth allocation given the prices and user association
equilibrium in the preceding section.
%Based on Theorem \ref{Thm:Price and User Association Equilibrium}, we can formulate the corresponding optimization problems, one for each of the mixed service and separate service scenarios, with the objective to maximize the total revenue from macro-cells and small-cells.
In the mixed service scenario, this is given by:
\begin{align}\label{Opt:Mixed Service}
\notag \text{maximize} \quad & S=(B_M+\lambda_SB_S)R_0u^{'}\Big(\frac{(B_M+\lambda_SB_S)R_0}{K_M+K_S}\Big)\\
\notag \text{subject to} \quad & B_M,B_S\ge 0, B_M+B_S\le B, B_S < B_{S,0}.
%\frac{\kappa N_fB_MR_0-N_mC_U}{\kappa N_m\lambda_SR_0}
\end{align}
where the objective is revenue, $K_M, K_S$ are given in (\ref{Eqn:Mixed Service}),
and $B_{S,0}$ is defined in \eqref{eq:BS0}.

In the separate service scenario, the objective becomes
\begin{equation}
S=B_MR_0u^{'}\Big(\frac{B_MR_0}{K_M}\Big) +
\lambda_SB_SR_0u^{'}       \Big(\frac{\lambda_SB_SR_0}{K_S}\Big)
\end{equation}
where $K_M, K_S$ are given in (\ref{Eqn:Separate Service}),
and the constraint $B_S \geq B_{S,0}$ applies.
%\begin{align}\label{Opt:Separate Service}
%\notag \text{maximize} \ &
%\notag \text{subject to} \ & B_M,B_S\ge 0, B_M+B_S\le B\\
%                        & B_S \ge \frac{\kappa N_fB_MR_0-N_mC_U}{\kappa N_m\lambda_SR_0}
%\end{align}

\begin{theorem}[Monopoly Bandwidth Allocation]\label{Thm:Optimal Bandwidth Allocation}
The optimal bandwidth allocation is unique and always corresponds to the separate service scenario.
%Moreover, all available bandwidth is used such that the split across macro- and small-cells  equalizes the marginal revenue increase with respect to per unit bandwidth allocation across both cells.
\end{theorem}

The optimal bandwidth allocation $(B_S^{\text{rev}}, B_M^{\text{rev}})$
can be determined by solving the necessary conditions given in Appendix \ref{Appen:Optimal Bandwidth Allocation}.
%For the class of $\alpha$-utility functions the optimal bandwidth allocation
%\begin{displaymath}\tag{P1} \label{Eqn:Revenue Opt with unlicensed}
%\left\{
%\begin{array}{ll}
%\Big[ \big( \frac{\kappa\lambda_SB_S^*R_0+C_U}{KN_f}   \big)^{-\alpha}
%-\alpha\big(\frac{\kappa\lambda_SB_S^*R_0+C_U}{KN_f}\big) ^{-\alpha}\\
%\quad \times \frac{\kappa\lambda_SB_S^*R_0}{\kappa\lambda_SB_S^*R_0+C_U}         \Big]  \lambda_S
% = (1-\alpha)\big( \frac{B_M^*R_0}{N_m}    \big)^{-\alpha} \\
% B_S^{*}+B_M^{*}=B
%\end{array}
%\right.
%\end{displaymath}
%\begin{align}
%\begin{split}
%& \frac{1-\alpha}{\lambda_S}\big( \frac{B_M^\text{rev}R_0}{N_m}    \big)^{-\alpha}  =  \Bigg[ \big( \frac{\kappa\lambda_SB_S^\text{rev}R_0+C_U}{\kappa N_f}   \big)^{-\alpha} \\
%& \qquad \quad -\alpha\big(\frac{\kappa\lambda_SB_S^\text{rev}R_0+C_U}{\kappa N_f}\big) ^{-\alpha} \frac{\kappa\lambda_SB_S^\text{rev}R_0}{\kappa\lambda_SB_S^\text{rev}R_0+C_U}         \Bigg] ,  \\
%& \qquad B_S^{\text{rev}}+B_M^{\text{rev}}  =B, \qquad B_S^\text{rev}, B_M^\text{rev} \geq 0.
%\end{split}
%\tag{P1} \label{Eqn:Revenue Opt with unlicensed}
%\end{align}
Although the solution
%\eqref{Eqn:Revenue Opt with unlicensed}
must be computed numerically, some general properties are easily established.
First, $B_M^\text{rev} > 0$ (always), but $B_S^\text{rev} > 0$
if and only if $C_U < C_U^{\text{rev}}$
(otherwise $B_S^\text{rev} =0$), where
\begin{equation} \label{Eqn:Threshold of C_U for revenue maximization}
C_U^{\text{rev}}=\frac{\kappa N_fBR_0}{N_m}\Big(  \frac{\lambda_S} {1-\alpha} \Big)^{\frac{1}{\alpha}}.
\end{equation}
When $\alpha \rightarrow 0^{+}$ or $\alpha \rightarrow 1^{-}$,
$C_U^{\text{rev}} \rightarrow +\infty$. That is, as the utility function
becomes either linear or logarithmic, the SP always allocates some bandwidth
to small-cells.

The optimal bandwidth allocation with no unlicensed spectrum
can be determined by setting $C_U=0$. %in \ref{Eqn:Revenue Opt with unlicensed}.
We will denote all associated quantities with a tilde, i.e.,
$(\tilde{B}_S^\text{rev}, \tilde{B}_M^\text{rev})$ is the optimized bandwidth
with $C_U=0$.
%In contrast, if there is no unlicensed spectrum, then the optimal bandwidth allocation $(\tilde{B}_S^\text{rev}, \tilde{B}_M^\text{rev})$ determined by \cite{Opt7-Chen13} yields:
%\begin{displaymath}\tag{P2} \label{Eqn:Revenue Opt without unlicensed}
%\left\{
%\begin{array}{ll}
%\lambda_S\big( \frac{\lambda_S\tilde{B}_S^\text{rev}R_0}{N_f}\big) ^{-\alpha}       =
%\big( \frac{\tilde{B}_M^\text{rev}R_0}{N_m}    \big)^{-\alpha} \\
% \tilde{B}_S^\text{rev}+\tilde{B}_M^\text{rev}=B
%\end{array}
%\right.
%\end{displaymath}
With no unlicensed spectrum, it is straightforward to show that
$\tilde{B}_S^\text{rev}=\tilde{\beta}B$ where
$\tilde{\beta} = \tfrac{N_f}{N_f+N_m \lambda_S^{1-1/\alpha}}$,
so that $\tilde{B}_S^{\text{rev}}>0$ and $\tilde{B}_M^{\text{rev}}>0$.
In contrast, with unlicensed spectrum the additional competition can cause
a revenue-maximizing SP to abandon small-cells altogether.
We will see a more drastic example of this effect when we consider competing SPs.

The next theorem compares the amount of bandwidth allocated to small-cells
with and without unlicensed spectrum.

%\begin{theorem}[Bandwidth allocation with/without unlicensed spectrum]
\begin{theorem}[Bandwidth Allocation Comparison]
\label{Thm:Comparison}
%Compared with the scenario without unlicensed spectrum, we have the following results: \\
There exists a threshold $C_U^{th}$ such that
%when the total rate capacity of WiFi network $C_U$ is less than $C_U^{th}$, the service provider should invest more bandwidth to small-cells when there is unlicensed spectrum. On the other hand, if $C_U$ is more than $C_U^{th}$, the service provider should invest less bandwidth to small-cells compared with the case without unlicensed spectrum. That is:
%\begin{align}
%\notag &\text{When } C_U < C_U^{th}, \quad B_S^\text{rev} > \tilde{B}_S^\text{rev}\\
%\notag &\text{When } C_U > C_U^{th}, \quad B_S^\text{rev} < \tilde{B}_S^\text{rev}
%\end{align}
when $C_U < C_U^{th}$, $B_S^\text{rev} > \tilde{B}_S^\text{rev}$,
and when $C_U > C_U^{th}$, $B_S^\text{rev} < \tilde{B}_S^\text{rev}$,
where \; $C_U^{th}=\beta^* \kappa \lambda_S\tilde{B}_S^\text{rev}R_0$
with $\beta^*$ being the unique strictly positive solution of
\begin{align}
%\notag &(1-\alpha)(\kappa\lambda_S\tilde{B}_S^{*}R_0)^{1+\alpha}=-C_U^{th}(\kappa\lambda_S\tilde{B}_S^*R_0)^{\alpha}+\\
%&(1-\alpha)(\kappa\lambda_S\tilde{B}_S^*R_0+C_U^{th})^{1+\alpha} \\
%& \beta=(1-\alpha) \Big( (1+\beta)^{1+\alpha} - 1\Big)
(1-\alpha) (1+\beta)^{1+\alpha} - \beta = 1-\alpha.
\end{align}
Furthermore, the SP's revenue is always less with unlicensed spectrum than without.
\end{theorem}

The last part of Theorem \ref{Thm:Comparison} is expected since unlicensed access
competes with the SP for fixed users.
The first part can be explained as follows. Compared to the case without unlicensed spectrum, when unlicensed spectrum is present if the SP keeps its bandwidth allocation the same, then it will clearly lose revenue as fewer users will use its small-cell service. To improve its revenue it could decrease $B_S$, shifting more resources to mobile users to make more revenue from them or it could increase $B_S$ to make its small-cell service more attractive and try to win back some users from the unlicensed network. However, increasing $B_S$ also results in a decrease in $p_S$ since the service rate per user increases (and a loss in revenue from the mobile users).  When $C_U$ is small, this decrease in $p_S$ is smaller since more users will switch to the small-cells making the second option more attractive. When $C_U$ is large enough, fewer users will switch to the small-cells per unit of additional bandwidth, making the first option more profitable.

%The change in marginal revenue per unit bandwidth increase in small-cells
%comes from two factors. On the one hand, this bandwidth increase
%leads to a larger capacity $C_S$. On the other hand, $p_S$ decreases
%since the average service rate increases. The increase in capacity dominates
%so that overall the revenue still increases.
%When the capacity of the unlicensed network $C_U$ is small, then
%compared with the scenario without unlicensed spectrum, $p_S$ decreases more slowly
%since the mass of fixed users in small-cells $K_S$ also increases when $B_S$ increases.
%In contrast, $K_S$ is fixed without unlicensed spectrum.
%Therefore the marginal revenue increase with $B_S$ is larger with unlicensed spectrum
%than without so the SP should allocate more bandwidth to small-cells.
%When $C_U$ becomes large, $p_S$ becomes small since most of the fixed users
%associate with the unlicensed network,
%so the SP obtains more marginal revenue by allocating
%more bandwidth to the macro-cells.

%In general, $\beta^*$ can only be computed numerically. However, for some specific choices of $\alpha$, we can solve $\beta$ in closed form. For example, when $\alpha=0.5$, $\beta^*=\tfrac{1+\sqrt{5}}{2}$ and $C_U^{th}=\tfrac{1+\sqrt{5}}{8}\lambda_S\tilde{B_S^{\text{rev}}}R_0$. The other cases in which we can analyze $C_U^{th}$ are the asymptotic scenarios when $\alpha$ approaches either 0 or 1.

When $\alpha\rightarrow 0^{+}$, $\kappa\rightarrow 0^+$ (so that $\alpha$-fair utility function becomes the linear utility function), $\beta^*\rightarrow 0^+$, and therefore, $C_U^{th}\rightarrow 0^+$. This implies that the SP should always invest less bandwidth to small-cells compared with the scenario without unlicensed spectrum. However, in this case the SP should still invest almost all bandwidth to small-cells, i.e., $B_S^{\text{rev}} \rightarrow B$. This is because $\alpha=0$ effectively corresponds to maximizing the sum rate. (This can be seen directly from the revenue function.)

%We can see this by analyzing the revenue function:
%\begin{align}
%\notag S&=B_MR_0p_M+\lambda_SB_SR_0p_S\\
%&=B_MR_0u^{'}\Big(\frac{B_MR_0}{K_M}\Big) +\lambda_SB_SR_0u^{'}       \Big(\frac{\lambda_SB_SR_0}{K_S}\Big)
%\end{align}
%\noindent When the $\alpha$-fair utility function becomes the linear utility function, the prices $p_M=u^{'}\Big(\frac{B_MR_0}{K_M}\Big)$, $p_S=u^{'}\Big(\frac{\lambda_SB_SR_0}{K_S}\Big)$ becomes constants and are equal. Therefore, in order to maximize the revenue, the SP only needs to maximize the total rate carried. As a result, the SP would allocate all bandwidth to small-cells. Note that, as expected, $\tilde{\beta}$ (the fraction of bandwidth assigned to small-cells when there is no unlicensed access) also coverages to $1$ as $\alpha\rightarrow 0^+$.

When $\alpha\rightarrow 1^{-}$, the utility function becomes logarithmic,
$\kappa \rightarrow \mathrm{e}^{-1}$, and $\beta^*\rightarrow \infty$,
so that $C_U^{th}\rightarrow \infty$. Hence, the SP should allocate
more bandwidth to small-cells with unlicensed spectrum.
Again,
in the limit the SP allocates all bandwidth to small-cells, i.e.,
$B_S^{\text{rev}} \rightarrow B$.
%$C_U^{th}\rightarrow +\infty$. That is to say, the SP should always invest more bandwidth to small-cells compared with the scenario without unlicensed spectrum. Actually in this case the service provider should also invest almost all bandwidth to small-cells.
We can see this by rewriting the revenue function as follows:
\begin{align}
\notag S&=B_MR_0u^{'}\Big(\frac{B_MR_0}{K_M}\Big) +\lambda_SB_SR_0u^{'}       \Big(\frac{\lambda_SB_SR_0}{K_S}\Big)\\
&=K_MR_Mu^{'}(R_M)+K_SR_Su^{'}(R_S).
\end{align}
\noindent For a log-utility function the revenue per mobile user
$R_Mu^{'}(R_M)$ and revenue per fixed user $R_Su^{'}(R_S)$ become constants
and are equal. Therefore maximizing revenue is equivalent to maximizing
$K_M$ and $K_S$. As a result, the SP allocates an arbitrarily small amount
of bandwidth to the macro-cells to guarantee that all mobile users are served,
and the remaining bandwidth to small-cells to maximize $K_S$.
In contrast, with no unlicensed access
$\tilde{\beta}$ converges to $\tfrac{N_f}{N_f+N_m} < 1$ and
$\lim_{\alpha\rightarrow 1^-}\tilde{B}_S^\text{rev}<B$.

\section{Social Welfare Maximization}\label{Sec:Social Welfare Maximization}
Now we change the objective function to social welfare
and analyze the corresponding prices, user association equilibrium
and bandwidth allocation.
%The following theorem summarizes the results.
\begin{theorem}[Social Welfare Maximization] \label{Thm:Social Welfare Maximization}
Given a single social welfare maximizing SP, in equilibrium we have the following properties:\\
{\it 1.} The prices and user association are the same as for revenue maximization
stated in Theorem \ref{Thm:Price and User Association Equilibrium}. \\
{\it 2.} The optimal bandwidth allocation is unique and corresponds to separate service. \\
%Moreover, all available bandwidth utilized such that the split equalizes the marginal utility increase of mobile users and fixed users with respect to the bandwidth in macro-cells and small-cells.\\
{\it 3.} Compared to the scenario without unlicensed spectrum,
%at the optimal bandwidth allocation,
with unlicensed spectrum the SP always allocates less bandwidth to small-cells
and more bandwidth to macro-cells.
\end{theorem}
\vspace{6pt}

%For the $\alpha$-utility functions we use in our model,
Similar with revenue maximization, the optimal bandwidth allocation
$(B_S^{\text{sw}}, B_M^{\text{sw}})$ can be determined
by solving the necessary conditions given in Appendix \ref{Appen:Social Welfare Maximization}.
%\begin{displaymath}\tag{P3} \label{Eqn:SW Opt with unlicensed}
%\left\{
%\begin{array}{ll}
%\frac{ (N_f)^{\alpha}\lambda_S\big((\kappa^{\alpha}+\kappa) C_U+ \kappa^{\alpha+1}\lambda_SB_S^{\text{sw}}R_0  \big)        }{ (\kappa\lambda_SB_S^{\text{sw}}R_0+C_U)^{\alpha+1}             } = (\frac{ B_M^{\text{sw}}R_0}{N_m  })^{-\alpha}, \\
%B_S^{\text{sw}}+B_M^{\text{sw}}=1, \qquad B_S^{\text{sw}}, B_M^{\text{sw}} \geq 0.
%\end{array}
%\right.
%\end{displaymath}
%We again observe some properties of the solution.
Also, it can be shown that $B_M^{\text{sw}} > 0$ (always),
and $B_S^{\text{sw}} > 0$ if and only if $C_U < C_U^{\text{sw}}$
(otherwise $B_S=0$), where
\begin{equation} \label{Eqn:Threshold of C_U for SW maximization}
C_U^{\text{sw}}=\frac{\kappa N_fBR_0\big[(\alpha+1)\lambda_S\big]^\frac{1}{\alpha}}{N_m}.
\end{equation}
As $\alpha \rightarrow 0$, $C_U^{\text{sw}} \rightarrow +\infty$, i.e.,
in the limiting case of a linear utility function,
the SP always allocates some bandwidth to small-cells,
even if $C_U$ is large. Similar with revenue maximization,
this is because this maximizes the total rate.
When $\alpha \rightarrow 1$,
$C_U^{\text{sw}} \rightarrow 2\mathrm{e}^{-1}\lambda_S \frac{N_fBR_0}{N_m}$
(as opposed to infinity for revenue maximization). With no unlicensed spectrum,
$\tilde{B}_S^{\text{sw}} > 0$ and $\tilde{B}_M^{\text{sw}} > 0$.
%can be determined simply by setting $C_U=0$ in \ref{Eqn:SW Opt with unlicensed}, which is the same as the revenue maximization without unlicensed spectrum, and therefore $\tilde{B}_S^{\text{sw}}, \tilde{B}_M^{\text{sw}}$ are always nonzero.
%If there is no unlicensed spectrum, then the optimal bandwidth allocation $(\tilde{B}_S^{\text{sw}}, \tilde{B}_M^{\text{sw}})$ can be determined by the following \cite{Opt7-Chen13}:
%\begin{displaymath}\tag{P4} \label{Eqn:SW Opt without unlicensed}
%\left\{
%\begin{array}{ll}
%\lambda_F\big( \frac{\lambda_S\tilde{B}_S^{\text{sw}}R_0}{N_f}\big) ^{-\alpha}       =
%\big( \frac{\tilde{B}_M^{\text{sw}}R_0}{N_m}    \big)^{-\alpha} \\
% \tilde{B}_S^{\text{sw}}+\tilde{B}_M^{\text{sw}}=B, \qquad \tilde{B}_S^{\text{sw}},\tilde{B}_M^{\text{sw}} \geq 0
%end{array}
%\right.
%\end{displaymath}
%which is the same as the revenue maximization without unlicensed spectrum, and therefore $\tilde{B}_S^{\text{sw}}, \tilde{B}_M^{\text{sw}}$ are always nonzero.

Part 3) in Theorem \ref{Thm:Social Welfare Maximization}
is due to the additional resources in the unlicensed network.
Since fixed users are better off with unlicensed spectrum,
to maximize the sum utility over all users, the SP allocates more bandwidth
to the macro-cells.

%In our previous work \cite{Opt7-Chen13} we showed that if there is no unlicensed spectrum,
We emphasize that with unlicensed spectrum,
Theorems \ref{Thm:Comparison} and \ref{Thm:Social Welfare Maximization} imply that
the optimal bandwidth allocation is different for revenue and social welfare maximization.
That is, the corresponding necessary conditions
%\ref{Eqn:Revenue Opt with unlicensed} and \ref{Eqn:SW Opt with unlicensed}
generally have different solutions. In contrast,
without unlicensed access, revenue and social welfare maximization
give the same bandwidth allocation
for $\alpha$-fair utility functions shown in our previous work\cite{Opt7-Chen13}.

\begin{figure}[htbp]
\centering
\includegraphics[width=0.45\textwidth,height=0.25\textwidth]{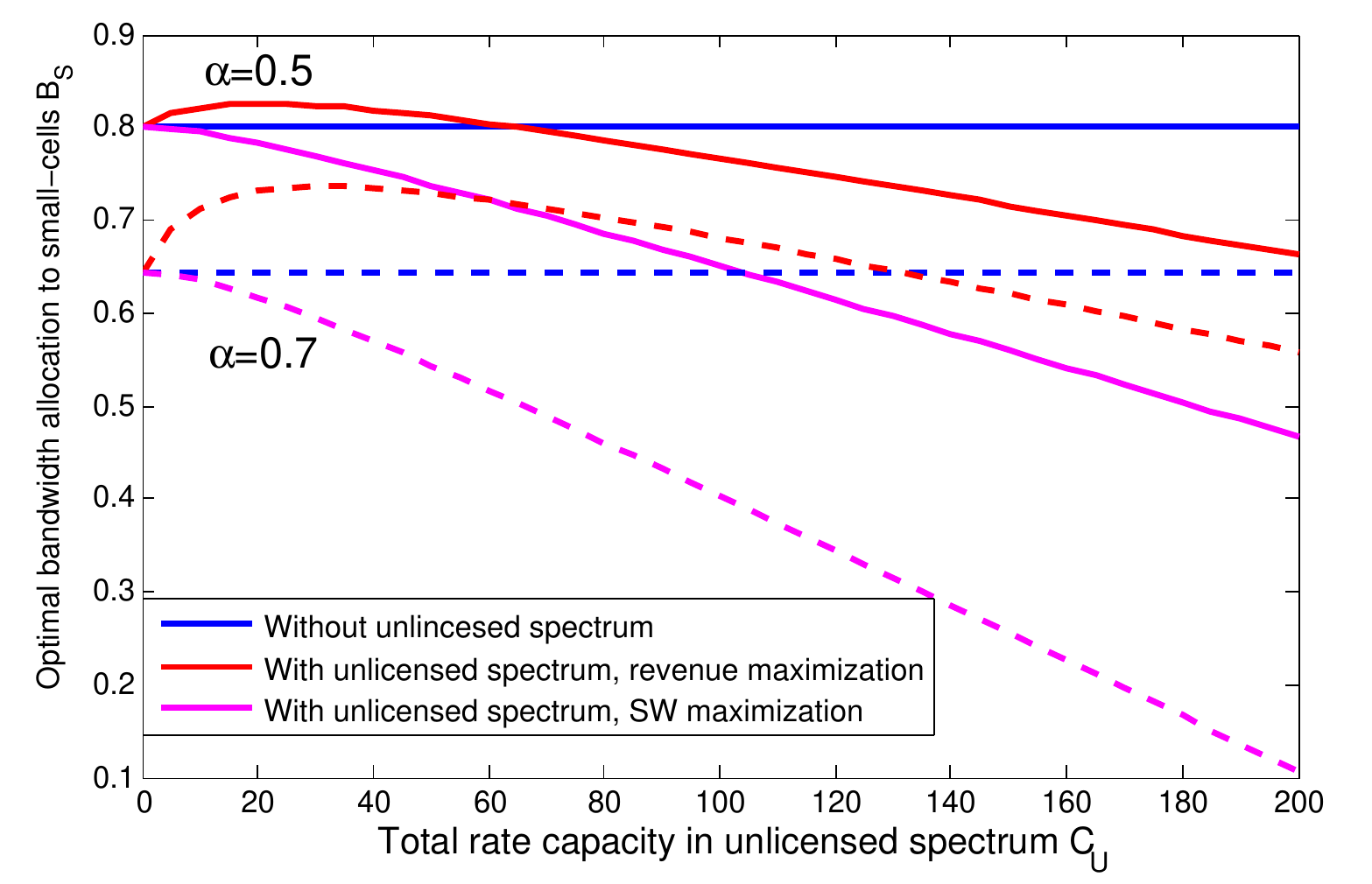}
\caption{Optimal bandwidth allocation to small-cells for a single SP
versus total unlicensed capacity.}
%($N_f=N_m=50,R_0=50,\lambda=4$). }
\label{Fig:Bandwidth_Monopoly_Revenue_SW}
\end{figure}

Figure \ref{Fig:Bandwidth_Monopoly_Revenue_SW} shows an example of
the optimal bandwidth allocation for different $\alpha$'s as the rate
offered by the unlicensed network increases.
Curves are shown for both revenue and social welfare maximization.
%We also compare with the case of no unlicensed access.
The system parameters are $N_f=N_m=50$, $R_0=50$, and $\lambda_S=4$.
For revenue maximization, the curve initially increases for small $C_U$,
and then decreases. In contrast, for social welfare maximization,
the curve is monotonically decreasing.

\section{Service Competition Among Multiple SPs}\label{Sec:Service Competition Among Multiple Service Providers}

In this section we study service competition among $N>1$ SPs
and investigate the corresponding sub-game perfect  equilibrium
with unlicensed spectrum.
%We now assume there are $N$ SPs operating in licensed spectrum and they have the same density of macro-/small-cells.
%Denote the set of SPs as $\mathcal{N}$. Denote the bandwidth allocation profile as $\mathbf{B}$ , total bandwidth allocation to small-cells and macro-cells as $B_S$, $B_M$, respectively.
%The user association rule is generally the same as the scenario with a single SP. The difference is that now mobile users also need to choose an SP to connect to its macro-cells. Fixed users also choose an SP to associate with its small-cells or they can choose the WiFi network in unlicensed spectrum.
Users choose the service (macro-/small-/unlicensed) which yields
the largest net payoff, and they fill the corresponding capacity accordingly.
In licensed spectrum, if multiple services offer the same price,
then the users are allocated across them in proportion
to the capacities. Once a particular service's capacity is exhausted,
then the leftover demand continues to fill the remaining services
in the same fashion. In unlicensed spectrum, fixed users always
get an average rate equal to the total rate divided
by the mass of fixed users associated with that network.

We again consider a sub-game perfect Nash equilibrium consisting of (i) A price equilibrium given a fixed bandwidth allocation; and (ii) An optimized
bandwidth allocation given that prices are set according to (i).
As for the scenario with a single SP, given a set of prices and bandwidth allocations,
the user association equilibrium
falls in one of two categories: a mixed service equilibrium in which
all macro-cells serve both mobile and a subset of fixed users,
and separate service equilibrium in which the macro-cells serve
only mobile users. The next theorem generalizes Theorem~\ref{Thm:Price and User Association Equilibrium} to multiple SPs.
%This is an equilibrium in which no SP can increase its revenue by unilaterally changing the price in its macro- or small-cells. It's very similar to the price choice that we analyzed in Section \ref{Sec:Price and User Association Equilibrium}, with the difference that now we have a game formulation and seek a price equilibrium. Again two service scenarios can arise, namely,\\
%1. Mixed service equilibrium: All macro-cells serve mobile users and a subset of fixed users jointly. Small-cells and WiFi network serve the remaining fixed users.
%For the mixed service equilibrium the prices $p_{i,M} = p_{i,S}$
%for all $i \in \mathcal{N}$, the set of SPs. For the
%separate service equilibrium the prices $p_{i,M} = p_{0,M}$
%and $p_{i,S} = p_{0,S}$ are each independent of $i$
%and $p_{0,M} > p_{0,S}$.
%across all macro- and small-cells for all SPs are the same, and this equilibrium price equalizes the total rate demand and total rate supply in licensed spectrum. \\
%2. Separate service equilibrium: All macro-cells only serve mobile users. All fixed users are served by small-cells and WiFi network. All macro-cell prices are equal and all small-cell prices are equal. The macro-cell price is higher than the small-cell price, and the prices equalize the total rate demand and rate supply in corresponding cells.

\begin{theorem}[Price Equilibrium with Multiple SPs]\label{Thm:Price NE}
Given fixed bandwidth allocations for all SPs,
there is a unique price equilibrium which clears the market. Further if
\[
\sum\limits_{i=1}^{N}B_{i, S} < \frac{\max\left(\kappa N_f\sum\limits_{i=1}^{N}B_{i, M}R_0-N_mC_U,0\right)}{\kappa N_m\lambda_SR_0}
\]
then a mixed service equilibrium holds. Otherwise, the  separate service equilibrium holds.

For the mixed service scenario, $p_{i,M} = p_{i,S}$ for each SP $i$; in the separate service case,
all SPs $i$ charge the same $p_{i,M}$ and $p_{i,S}$, with $p_{i,M} > p_{i,S}$.
\end{theorem}
\vspace{5pt}

%Using the characterization of the price equilibrium,
The next theorem characterizes the equilibrium for
the bandwidth allocation stage and thus the overall sub-game perfect Nash equilibrium for the game.
Before stating this we define the following types of
equilibria: A  \emph{Macro-Small Nash Equilibrium (MSNE)} is one where all SPs allocate bandwidth to both macro- and small-cells.  A \emph{Macro-Preferred Nash Equilibrium (MPNE)} is one where some SPs allocate bandwidth to both macro- and small-cells and
the remaining SPs allocate bandwidth to macro-cells only. Finally, a \emph{Macro-only Nash Equilibrium (MNE)} is one where all SPs allocate bandwidth to macro-cells only.

\begin{theorem}[Nash Equilibrium]\label{Thm:Existence and Uniqueness of NE}
There always exists a unique sub-game perfect Nash equilibrium
and it corresponds to the separate service scenario.
In the equilibrium fixed users in small-cells achieve
a higher average rate than mobile users in macro-cells.
Moreover, this Nash equilibrium can only be one of the following types:
an MSNE, an MPNE or an MNE.
\end{theorem}

\begin{figure}[htbp]
\centering
\includegraphics[width=0.40\textwidth,height=0.25\textwidth]{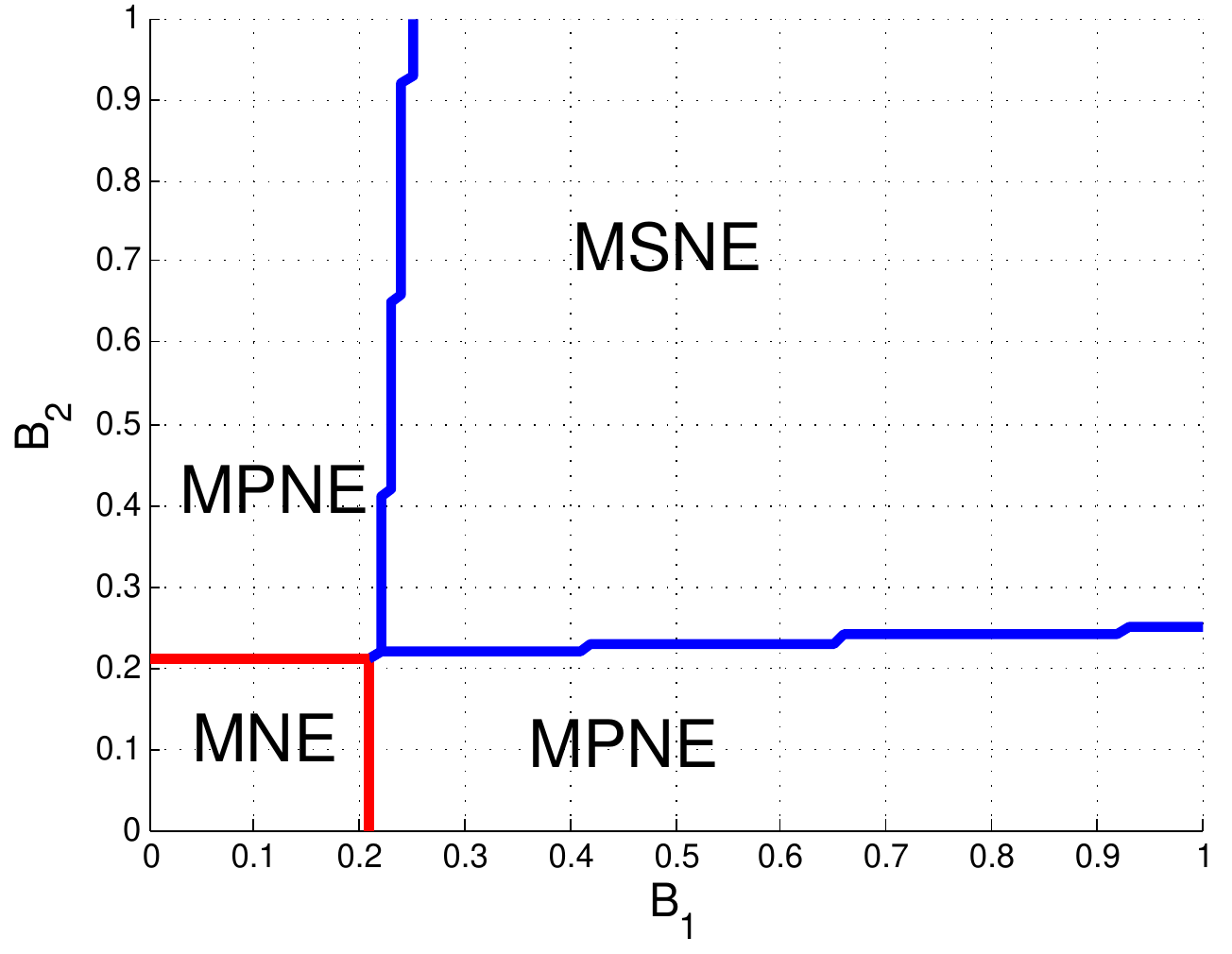}
\caption{Nash equilibrium regions for 2 SPs.
The system parameters are $\alpha=0.5, N_m=N_f=50, R_0=50, \lambda_S=4, \lambda_W=3, B_U=1$.}
\label{Fig:NE_Region_2SP}
\end{figure}

%\noindent The following remarks are in order:\\
If there is no unlicensed spectrum, then for
the $\alpha$-fair utility function only an MSNE
exists and it is always efficient, i.e., maximizes social welfare~\cite{Competition5-Chen15}.
Here, the presence of unlicensed access can cause
a subset of the SPs to engage in macro-service only.
Further, for any number of SPs, it can be shown that, in general,
none of the equilibrium categories (including MSNE) are efficient.
%and serve mobile users.\\
%3. With unlicensed access, for certain parameter settings we can also have an MNE, which is an even more extreme form of an equilibrium wherein all SPs only activate their macro-cells and only serve mobile users. All the fixed users in this equilibrium are served by the WiFi network.
Figure~\ref{Fig:NE_Region_2SP} illustrates the Nash equilibrium regions for two SPs
as a function of the available bandwidths $B_1$ and $B_2$.
When $B_1$ and $B_2$ are sufficiently large, then the equilibrium
is an MSNE, whereas if $B_1$ and/or $B_2$ become sufficiently small,
the equilibrium transitions so that at least one SP serves only mobile users.

%An MSNE (if it exists) has nice analytical properties that are summarized in the following proposition.
\begin{proposition}[MSNE Properties]
Assuming an MSNE, for any two SPs $i$ and $j$
with total bandwidth $B_i$ and $B_j$, the following properties hold:\\
1) \emph{Symmetry}: If $B_i=B_j$, then each SP's bandwidth allocation
must be the same, i.e., $B_{i, S}=B_{j, S}, B_{i, M}=B_{j, M}$.\\
2) \emph{Monotonicity}: If $B_i>B_j$, then SP $i$ allocates more bandwidth
to both macro- and small-cells than SP $j$, i.e., $B_{i, S}>B_{j, S}, B_{i, M}>B_{j, M}$.
\end{proposition}

%{\color{red}
%In addition, at an MSNE we can further compare the total bandwidth allocation to small-cells by all SPs in scenarios with and without unlicensed spectrum. Denoting $\tilde{B}_S$ as the sum total bandwidth allocation in small-cells by all SPs without unlicensed spectrum, we have the following proposition.

%Given a set of system parameters, if it results in an MSNE in the scenario with unlicensed access, then with the same parameters in the scenario without unlicensed access we also have an MSNE.
Denote the total bandwidth allocated to small-cells by all SPs
with and without unlicensed spectrum as $B_S$ and $\tilde{B_S}$, respectively.
%The following proposition characterizes the comparison between $B_S$ and $\tilde{B_S}$.

\begin{proposition}[MSNE Bandwidth Comparison] \label{Prop:MSNE Bandwidth Comparison}
If an MSNE exists with unlicensed access, then
%with the same system parameters an MSNE also exists without unlicensed access.
%Moreover,
%the total bandwidth allocated to small-cells in the scenario
%with unlicensed access is always less than that of without unlicensed access,
$B_S<\tilde{B_S}$.
\end{proposition}

Hence, for the MSNE case, competing SPs reduce the bandwidth allocation to small-cells
when unlicensed spectrum is introduced.

%In Section \ref{Sec:Optimal Bandwidth Allocation} it is shown that for the monopoly case, the comparison of the bandwidth allocation to small-cells between the two scenarios depends on the amount of total rate capacity in unlicensed spectrum. {\color{red} However, here with multiple SPs, at MSNE it becomes independent though.}

%Compared with MSNE or MPNE, MNE is an extreme type of Nash equilibrium since all SPs totally abandon small-cells and only invest bandwidth in macro-cells. The following proposition characterizes the sufficient and necessary condition for the existence of MNE:
\begin{proposition}[MNE Conditions]\label{Prop:Condition for MNE}
An MNE holds if and only if
\begin{equation}\label{Eqn:Condition for MNE}
C_U \geq \frac{R_0 \sum_{i=1}^N B_i}{\big(1-\alpha\frac{B_{\max}}{\sum_{i=1}^N B_i}\big)^{\frac{1}{\alpha}}} \frac{ \kappa N_f \lambda_S^{\frac{1}{\alpha}}}{N_m},
%\lambda_S\big(  \frac{C_U}{\kappa N_f} \big)^{-\alpha} \le \big( \frac{\sum\limits_{i=1}^{N}B_iR_0 }{N_m}\big)^{-\alpha} -\alpha \frac{B_{i_{\max}}R_0}{N_m} \big( \frac{\sum\limits_{i=1}^{N}B_iR_0 }{N_m}\big)^{-\alpha-1},
\end{equation}
where $B_{\max}=\max\{B_i: i \in \mathcal{N}\}$.
\end{proposition}

For $N=1$ (monopoly) this condition yields the threshold in
(\ref{Eqn:Threshold of C_U for revenue maximization}) so that
the SP allocates no bandwidth to small-cells.

\section{Unlicensed Bandwidth and Social Welfare}
\label{Sec:Licensed vs Unlicensed Spectrum}
%Thus far, we've compared the bandwidth allocation and social welfare in both scenarios with and without unlicensed spectrum.
%In the analysis thus far, we've characterized the existence of equilibria given a bandwidth assignment in terms of licensed and unlicensed spectrum.
In this section we study the effect of increasing unlicensed bandwidth on social welfare.
Using the preceding framework, we can determine
%and the optimal assignment of spectrum to licensed or unlicensed bands in terms of maximizing social welfare. In particular, we will determine
the specific mix of unlicensed/licensed spectrum such that the market equilibrium
yields the same social welfare as that achieved by a social planner.
%, i.e., the conditions for the price of anarchy to be 1.
This is motivated by the scenario in which a spectrum regulator,
such as the FCC, must determine how much newly available spectrum
will be licensed or unlicensed. We assume a total available bandwidth $B$
and consider the following scenarios:

1) {\em Efficient allocation}: A social planner determines the bandwidth allocation
to macro-cells $B_M$, small-cells $B_S$, and unlicensed network $B_U$ that
maximizes total utility. We will denote the optimal allocation as
$B_M^{\text{opt}}, B_S^{\text{opt}}, B_U^{\text{opt}}$,
and use the corresponding social welfare as a benchmark.

2) {\em Market equilibrium}: Here a social planner determines the
bandwidth assigned to licensed spectrum $B_L$ and unlicensed spectrum $B_U$.
Each of the $N$ SPs operating in licensed spectrum obtains the same portion
of total bandwidth $B_{i}\equiv\tfrac{B_L}{N}$, and then further determines
the split of $B_{i}$ between $B_{i_M}$ and $B_{i_S}$ to maximize its revenue.
This scenario corresponds to the more practical setting in which
the regulator sets aside part of the available bandwidth as unlicensed,
and grants licenses for the remainder.
%spectrum allocation process with a monopoly service provider ($N=1$)
%or competitive service providers
%($N\ge 2$ including the limiting regime of $N\rightarrow \infty$).
We will denote the bandwidths that maximize social welfare in this scenario
as $B_L^{*}, B_U^{*}$.

In the first scenario the social planner determines the bandwidth assignment
without explicit pricing.
%so that small-cells and unlicensed access are essentially the same since both
%only serve fixed users.
The optimal bandwidth allocation equalizes the marginal utility for mobile
and fixed users. It is easy to verify that the efficient allocation corresponds
to separate service.
% so that macro-cells only serve mobile users and fixed users only associate with small-cells or unlicensed access.
In the market equilibrium scenario, we will use the results
from Sections \ref{Sec:Optimal Bandwidth Allocation} and
\ref{Sec:Service Competition Among Multiple Service Providers}
to determine the optimal bandwidth assignments. We will also present social welfare curves for the asymptotic regime of many SPs, i.e., as $N\rightarrow\infty$, for which we will use the characterization below (full version in Appendix \ref{Appen:MSNE Conditions}).
%Before presenting the results we first present a characterization of the asymptotic regime as $N\rightarrow\infty$
%using Corollary \ref{Cor:SymmetricCase}
%and setting $C_U=B_U \lambda_U R_0$.

\begin{proposition}\label{Prop:SymmetricAsymptoticCase}
If we have
\begin{align}
B_U \lambda_U > B \frac{\kappa N_f \lambda_S^{\frac{1}{\alpha}}}{N_m},
\end{align}
then there exists an $N^*(B_U \lambda_U)$ such that for all
$N > N^*(B_U \lambda_U)$ we have an MNE.
%so that $B_{i,M}=B/N$ and
%($B_{i,S}=0$ for all $i\in \mathcal{N}$).
Otherwise, we always have an MSNE ($B_{i,M} > 0$ and $B_{i,S} > 0$ $\forall$ $i\in \mathcal{N}$,
and $\forall$ $N$).
%with $\lim_{N\rightarrow \infty} (N B_{i,M}, N B_{i,S})=(B_M, B_S)$ for all $i$ for some $B_M > 0$ and $B_S \geq 0$ with $B_M+B_S=B$.
%with $\lim_{N\rightarrow \infty} (N B_{i,M}, N B_{i,S})=(B_M, B_S)$ for all $i$ with
%\begin{align}
%B_S =  \frac{B- \frac{B_U \lambda_U N_m}{\kappa N_f \lambda_S^{\frac{1}{\alpha}} }}{1+\frac{\lambda_S N_m}{\lambda_S^{\frac{1}{\alpha}} N_f}}, B_M =  \frac{\bigg(B+B_U \frac{\lambda_U}{\kappa \lambda_S} \bigg) \frac{\lambda_S N_m}{\lambda_S^{\frac{1}{\alpha}} N_f}}{1+\frac{\lambda_S N_m}{\lambda_S^{\frac{1}{\alpha}} N_f}}.
%\end{align}
%Furthermore,
%\begin{align}
%\begin{split}
%& K_S=\frac{\kappa B_S \lambda_S N_f}{\kappa B_S \lambda_S+B_U\lambda_U}, K_U=\frac{B_U\lambda_U N_f}{\kappa B_S \lambda_S+B_U\lambda_U}, \\ &
%K_M=N_m, R_S = \frac{R_0(\kappa B_S \lambda_S+B_U\lambda_U)}{\kappa N_f},\\
%&  R_M=\frac{B_M R_0}{N_m}, R_U = \frac{R_0(\kappa B_S \lambda_S+B_U\lambda_U)}{N_f}.
%\end{split}
%\end{align}
\end{proposition}

Given newly available bandwidth $B$, the optimal split into licensed and unlicensed
subbands depends on
%The results of the social welfare maximizing bandwidth assignment under either the ideal case or the market equilibrium depend on
the relative values of $\lambda_S$ and $\lambda_U$ (proofs in Appendix \ref{Appen:Licensed vs Unlicensed Spectrum}). We have the following cases:\\
\noindent a) $\lambda_S>\lambda_U$: In this case an efficient allocation
by a social planner would assign all spectrum to macro- and small-cells,
i.e., there is no unlicensed network.
The optimal bandwidth assignment is then:
\begin{equation}\label{Eqn:BWSplit1}
B_M^{\text{opt}}=\frac{N_mB}{N_m+\mu_S N_f}, B_S^{\text{opt}}=\frac{\mu_S N_fB}{N_m+\mu_S N_f}, B_U^{\text{opt}}=0,\\
\end{equation}
where $\mu_S:=\lambda_S^{\frac{1}{\alpha}-1}$.
This is also true for the market equilibrium since it is shown in
\cite{Opt7-Chen13,Competition5-Chen15} that without unlicensed spectrum,
maximizing revenue is the same as maximizing social welfare
for $\alpha$-fair utility functions, independent of $N$.
%\begin{equation}
%B_L^{*}=B, B_U^{*}=0,
%\end{equation}
%with the total bandwidth allocation between macro- and small-cells exactly given by \eqref{Eqn:BWSplit1}. Furthermore, since $\lambda_U<\lambda_S$, we can easily show that any other assignment leads to a decrease in social welfare so that this is the unique bandwidth assignment.
Hence in this case the market equilibrium achieves the efficient allocation.

\noindent b) $\lambda_S=\lambda_U$:
%so that there is no difference between small-cells and unlicensed access. Therefore in the ideal case
In this case the social planner only needs to consider the bandwidth assigned
to macro-cells; the bandwidth split between small-cells and unlicensed access
can be arbitrary. Here the optimal assignment satisfies
\begin{equation}
B_M^{\text{opt}}=\frac{N_mB}{N_m+\mu_S N_f}, B_S^{\text{opt}}+B_U^{\text{opt}}=\frac{\mu_S N_fB }{N_m+\mu_S N_f},
\end{equation}
where $\mu_S=\lambda^{\frac{1}{\alpha}-1}$.
This includes the two extremes
$(B_S^{\text{opt}}>0, B_U^{\text{opt}}=0)$ and
$(B_S^{\text{opt}}=0, B_U^{\text{opt}}>0)$.
Hence for the market equilibrium case, an optimal bandwidth assignment strategy
is to allocate all bandwidth as licensed.
As in case a), this achieves the efficient allocation.
%using the results of \cite{Opt7-Chen13,Competition5-Chen15} the price of anarchy is 1.
However, here there may exist another efficient allocation in which
the fixed users are served by the unlicensed network.
For the competing SPs, that corresponds to an MNE, i.e., the SPs
only allocate their licensed bandwidth to the macro-cells
($B_{i,S} =0$ for all $i\in \mathcal{N}$).

%The following proposition provides conditions into whether or not
%such an efficient allocation exists.

%where once again the price of anarchy is 1. At this second optimal point, since the utility value is fixed, mobile and fixed users should have the same average rate as the first optimal point with $B_U=0$ but fixed users are served by unlicensed access only. In other words, all SP(s) only allocate bandwidth to macro-cells (MNE in the competitive scenario).

%As long as we can guarantee this, it would achieve the same optimal social welfare.
The bandwidth allocation corresponding to this second optimal assignment, along
with the condition for its existence, is
%strategy and corresponding condition that guarantees its optimality is given as follows:
\begin{align*}
%&B_L^{*}=B, B_U^{*}=0,  \\
%\text{or } &
& B_L^{*}=\frac{ N_mB }{N_m+\mu_S N_f}, B_U^{*}=\frac{\mu_S N_fB }{N_m+\mu_S N_f}, \\
%\text{if }&\kappa^{\alpha}+\frac{\alpha}{N} \le 1 \text{ holds.}
& \qquad \text{if } N\ge 2 \text{ or } N=1 \text{ and } 0<\alpha \le 0.5.
\end{align*}
For $N=1$, if $\alpha \in (0.5, 1)$, then this second optimal point does not exist
and the unique optimal bandwidth allocation corresponds to no unlicensed spectrum.
All other bandwidth assignments yield lower social welfare.

\noindent c) $\lambda_S<\lambda_U$: In this case a social planner
assigns spectrum to the macro-cell and unlicensed networks only;
there is no small-cell network.
The optimal allocation is:
\begin{equation}
B_M^{\text{opt}}=\frac{N_mB}{N_m+\mu_U N_f}, B_S^{\text{opt}}=0, B_U^{\text{opt}}=\frac{\mu_U N_fB}{N_m+\mu_U N_f},\\
\end{equation}
where $\mu_U=\lambda_U^{\frac{1}{\alpha}-1}$.
For the market equilibrium, allocating all bandwidth to licensed spectrum
no longer achieves the efficient allocation. The only possibility for achieving
an efficient allocation, then, is to allocate $B_U^{\text{opt}}$
to unlicensed access. This achieves the efficient allocation
if and only if the corresponding equilibrium is an MNE.
The corresponding bandwidth assignment, along with the condition
that guarantees its existence are:
\begin{align*}
&B_L^{*}=\frac{N_mB}{N_m+\mu_U N_f}, B_U^{*}=\frac{\mu_U N_fB}{N_m+\mu_U N_f}, \\
%\text{if }&\kappa^{\alpha}\frac{\lambda_S}{\lambda_U}+\frac{\alpha}{N} \le 1 \text{ holds.}
& \qquad \text{if } N\ge 2 \text{ or } N=1 \text{ and } 0<\alpha \le \alpha_0,
\end{align*}
where $0<\alpha_0<1$ is the unique solution to
$\kappa^{\alpha}\frac{\lambda_S}{\lambda_U}+\alpha = 1$.
All other bandwidth assignments yield lower social welfare.
Also, for $N=1$, if $\alpha \in (\alpha_0, 1)$, then
%the optimal bandwidth assignment is hard to characterize analytically
the optimal bandwidth allocation is not efficient.\footnote{The market equilibrium
does achieve the efficient allocation when $B_U=0$ \cite{Opt7-Chen13,Competition5-Chen15}.
Here we have that for the efficient allocation $B_U>0$.}
%However, from the ideal case, we know that optimal social welfare is achieved for some positive value of $B_U$.

To summarize, in contrast to the results of
\cite{Opt7-Chen13,Competition5-Chen15}, with the addition of unlicensed spectrum
%only under a specific bandwidth assignment(s) (common to all market scenarios) is
it is possible to achieve efficiency only with a specific split of licensed
and unlicened spectrum, even as $N\rightarrow \infty$ (perfect competition).
Moreover, when $\lambda_S < \lambda_U$, the optimal bandwidth assignment
(when it exists) is an MNE.

%We then present three figures illustrating the results above. Here we assume the total bandwidth is $B=2$ and the other parameters are $\alpha=0.5, N_f=N_m=50$, $R_0=50$, $\lambda_S=4$. The three figures (Figure \ref{Fig:Plot_4cases_2}, \ref{Fig:Plot_4cases_1}, \ref{Fig:Plot_4cases_3} each corresponding to the cases above) plot the social welfare achieved in four different scenarios, i.e, benchmark case, monopoly SP, two competitive SPs and infinite competitive SPs. Note that in the figures the straight benchmark line shows the optimal social welfare, while for the other three scenarios the utility is plotted as a function of $B_U \in [0,B]$. We can see that when $\lambda_S >\lambda_U$, the unique optimal bandwidth allocation strategy is to allocate all bandwidth to licensed spectrum.  Because the two conditions both hold for $\alpha=0.5$, when $\lambda_S= \lambda_U$, we have two different bandwidth assignments for both monopoly and competitive scenarios that can achieve the optimal benchmark social welfare. While when $\lambda_S < \lambda_U$, since $\alpha=0.5<\alpha_0\approx0.638$ (for $\tfrac{\lambda_S}{\lambda_U}=0.8$) we have a unique such bandwidth assignment.

\begin{figure}[htbp]
\centering
\includegraphics[width=0.45\textwidth,height=0.25\textwidth]{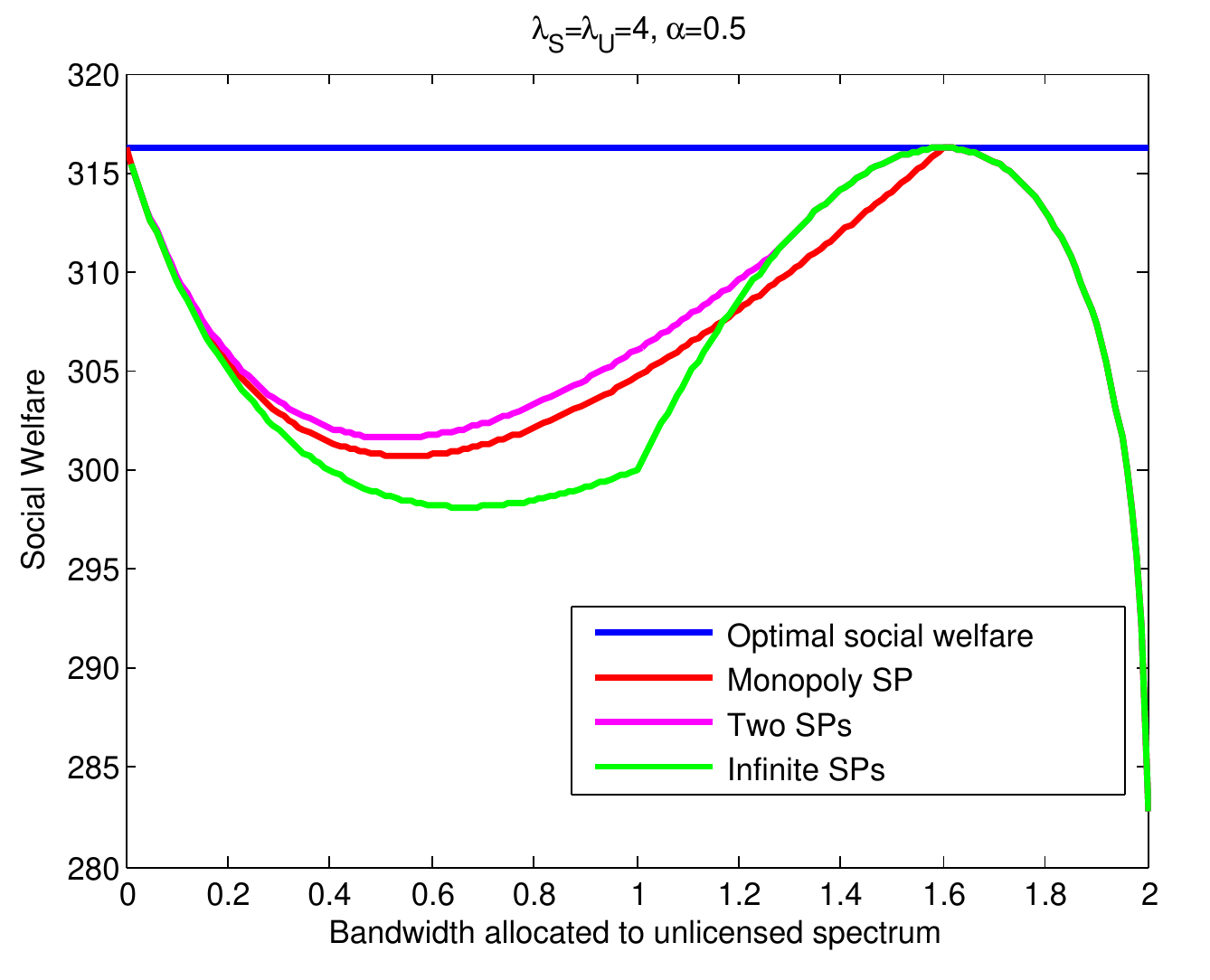}
\caption{Social welfare versus unlicensed bandwidth with $\lambda_S= \lambda_U$.}
%($N_f=N_m=50,R_0=50,\lambda=4$). }
\label{Fig:Plot_4cases_1}
\end{figure}

\begin{figure}[htbp]
\centering
\includegraphics[width=0.45\textwidth,height=0.25\textwidth]{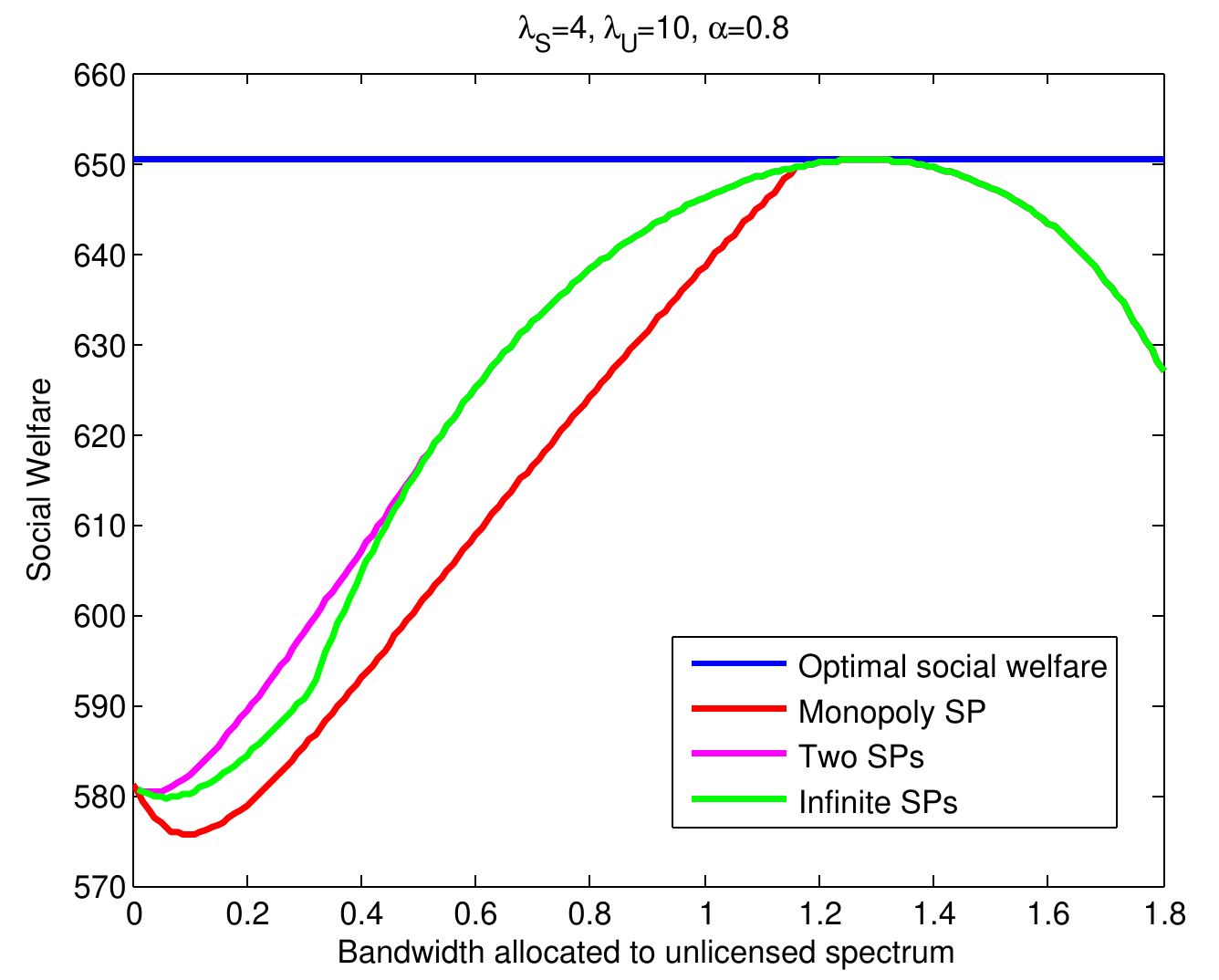}
\caption{Social welfare versus $B_U$ with $\lambda_S< \lambda_U$.
Here a monopolist can be efficient.}
%($N_f=N_m=50,R_0=50,\lambda=4$). }
\label{Fig:Plot_4cases_3}
\end{figure}

\begin{figure}[htbp]
\centering
\includegraphics[width=0.45\textwidth,height=0.2\textwidth]{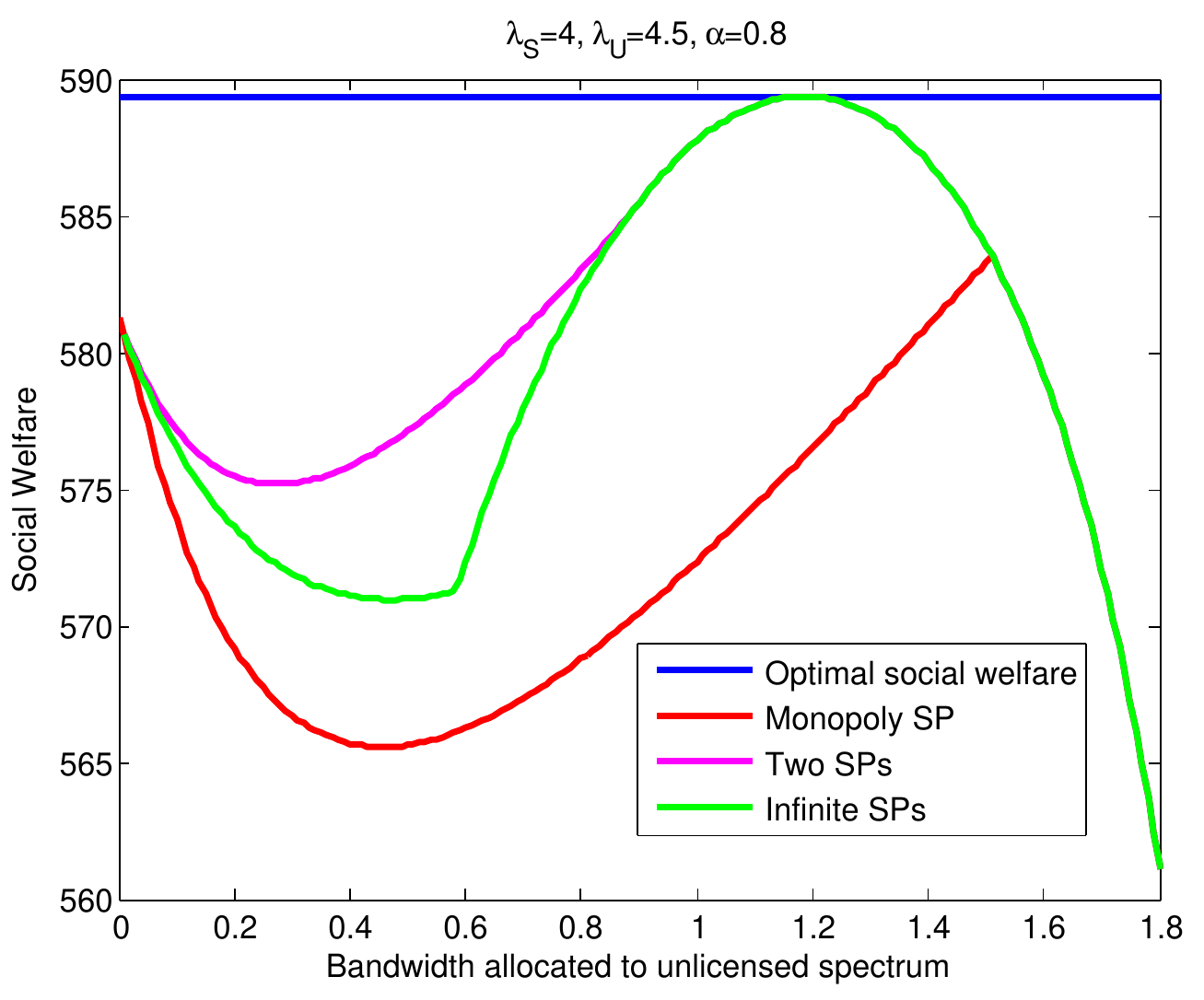}
\caption{Social welfare versus $B_U$ with $\lambda_S< \lambda_U$. Here the monopolist
is always inefficient.}
%($N_f=N_m=50,R_0=50,\lambda=4$). }
\label{Fig:Plot_4cases_4}
\end{figure}

Figures \ref{Fig:Plot_4cases_1}-\ref{Fig:Plot_4cases_4} illustrate the preceding observations.
They show social welfare versus the amount of unlicensed bandwidth
for total bandwidth $B=2$, $N_f=N_m=50$, $R_0=50$, and $\lambda_S=4$.
Each figure shows four plots corresponding to the efficient allocation (straight line),
monopoly SP, two competitive SPs, and perfect competition.
%Note that in the figures the straight benchmark line shows the optimal social welfare, while for the other three scenarios the utility is plotted as a function of $B_U \in [0,B]$.

Figure \ref{Fig:Plot_4cases_1} illustrates case b), where
$\alpha=0.5$ is chosen so that two different bandwidth assignments give
the efficient allocation for all scenarios.
We also compute $B_U^{\text{opt}}=1.6$, $\tfrac{C^{\text{sw}}_U}{\lambda_U R_0}=:B^{\text{sw}}_U\approx1.39$, and $\tfrac{C^{\text{rev}}_U}{\lambda_U R_0}=:B^{\text{rev}}_U=1.6$. A social welfare maximizing monopolist would start ignoring small-cells when $B_U\approx1.39$, while a revenue maximizing monopolist takes the same action exactly at the value of $B_U$ where social welfare is maximized.

Figure \ref{Fig:Plot_4cases_3} illustrates case c) where a monopolist
can be efficient ($\tfrac{\lambda_S}{\lambda_U}=0.4, \alpha=0.8$).
Here $B^{\text{rev}}_U\approx 1.16$, $B^{\text{opt}}_U\approx1.28$ and $B^{\text{sw}}_U\approx 0.56$, so that both a social welfare or revenue maximizing monopolist abandon small-cells for small enough values of $B_U$. Figure \ref{Fig:Plot_4cases_4} illustrates case c)
where a monopolist is not efficient
($\tfrac{\lambda_S}{\lambda_U}=\tfrac{8}{9}, \alpha=0.8$).
Here $B^{\text{rev}}_U\approx 1.51$, $B^{\text{opt}}_U\approx1.19$ and $B^{\text{sw}}_U\approx 0.92$. Note that the revenue maximization objective makes the monopolist abandon small-cells only for large $B_U$, well after $B_{U}^{\text{opt}}$.

We do not plot an example for case a) as it is easily understood.
When $\lambda_S>\lambda_U$, all scenarios achieve the maximum social welfare
when all bandwidth is allocated to licensed spectrum.
This is because when $B_U=0$, from \cite{Opt7-Chen13, Competition5-Chen15} maximizing revenue is the same as maximizing
social welfare for any $N$.

We make several observations from the figures.
First, the three curves corresponding to the market scenarios ($N=1,2$, and $N \to \infty$)
all have a ``kink'', or turning point after which the curves are concave
with $B_U$.
This corresponds to the transition from an MSNE to MNE (macro-cells only).
%allocate bandwidth to small-cells, i.e, result in an MNE in competitive scenario.
That is, to the right of the turning point bandwidth is allocated only to
macro-cells and unlicensed access, and the social welfare is always concave
in $B_U$. For the MNE the social welfare
is the same for the three market scenarios, so that
the three curves overlap in this region, i.e., when $B_U$
becomes sufficiently large. This common strictly concave function
gives the sum utility as function of $B_U$ if there are no small-cells.
It can be extended to all $B_U \in (0,B)$ and has a unique maximizer in $(0,B)$. %Owing to this in Figures \ref{Fig:Plot_4cases_3} and \ref{Fig:Plot_4cases_4} we do not display the entire range of $B_U$ as the social welfare only decreases from that point onwards.

When $\lambda_S=\lambda_U$, our results show that for $\alpha=0.5$
we have two optimal points. Therefore in Fig. \ref{Fig:Plot_4cases_1}
the curves first decrease and then increase to the second optimal point
as part of the concave function we previously described.
When $\lambda_S<\lambda_U$, with a small amount of bandwidth allocated to unlicensed spectrum,
we obtain slightly more rate, but this does not maximize social welfare,
and so the social welfare decreases. However, when we increase $B_U$,
the fixed users' utility increases with rate, and this effect dominates
even though we are not allocating the rate efficiently.
As a result, the social welfare goes up again.
As $B_U$ approaches $B$, mobile users in macro-cells suffer and therefore
the social welfare decreases again.

%\begin{figure}[htbp]
%\centering
%\includegraphics[width=0.5\textwidth,height=0.25\textwidth]{Plot_4cases_2.pdf}
%\caption{Social welfare performance in monopoly and competitive scenarios with bandwidth allocation to unlicensed spectrum when $\lambda_S> \lambda_U$.}
%\label{Fig:Plot_4cases_2}
%\end{figure}

\section{Conclusion}\label{Sec:Conclusion}
We have presented a model for allocating bandwidth in a HetNet with
both licensed and unlicensed spectrum, taking into account 
the pricing strategies of the SPs.
Our results characterize the equilibrium allocations assuming
that the SPs maximize revenue or social welfare.
For a monopoly SP that maximizes revenue, we show that
the presence of a small amount of unlicensed spectrum may cause the SP to
allocate more bandwidth to its competing small-cell network.
However, when maximizing social welfare the SP always allocates
less bandwidth with unlicensed spectrum.
With multiple competing SPs the (unique) equilibrium is one
of three different types, depending on the system parameters,
%In this paper we studied the pricing and bandwidth allocation problem 
%in heterogeneous wireless networks with both licensed and unlicensed 
%spectrum in two different scenarios: monopoly and competitive.  
%In both scenarios we compared the optimal bandwidth allocation 
%with the scenario without unlicensed spectrum. 
%We showed that in the monopoly scenario, 
%for revenue maximization whether the service provider 
%should invest more or less bandwidth to small-cells 
%depends on the rate in the unlicensed spectrum. 
%On the other hand, for social welfare maximization 
%the service provider should always invest less bandwidth 
%in small-cells compared the case without unlicensed spectrum. 
%In the competitive scenario, we showed that a 
%%unique Nash equilibrium always exists. 
%However, depending on different system parameters, 
including one in which the SPs do not allocate any bandwidth
to a small-cell network (e.g., when the available bandwidth
is small). We observe that in general, these
equilibria do not achieve an efficient allocation
corresponding to the maximum social welfare
even when the number of competing SPs is large.
In contrast, without unlicensed spectrum the equilibrium
is always efficient for the class of $\alpha$-fair utility
functions considered here.

We have used this framework to analyze the effect of 
unlicensed bandwidth on social welfare.
If the small-cell network offers higher spectral efficiency
than the unlicensed network, then according to our model,
allocating all of the bandwidth as licensed is efficient.
Otherwise, if the unlicensed network offers higher spectral efficiency,
then we observe that there is a unique mix of unlicensed and
licensed spectrum that maximizes social welfare, but that
mix may or may not be efficient when the licensed spectrum 
is allocated to revenue-maximizing SPs.

In practice, a newly deployed licensed small-cell network may be based 
on LTE or WiFi, and the unlicensed (open access) network may be based
on WiFi or LTE-U. Given the same density of access points,
a managed LTE network would likely
have the highest spectral efficiency, in which case, according to our model,
allocating bands as licensed will maximize the social welfare.
Of course, this assumes that the utility of a band depends only on the offered rate.
Advocates of unlicensed spectrum have pointed to other properties,
which are not taken into account here, such as lowering barriers
to entry and the potential for developing new technologies 
and business models.\footnote{See, for example, \cite{MilgromLevinEilat2011}.
The relative benefits of unlicensed versus licensed spectrum
remain controversial, e.g., see \cite{Hazlett2010}.}
Also not taken into account here is investment, which may
pose different types of barriers to entry in licensed versus
unlicensed spectrum. Incorporating investment within the framework
presented here is left for future work.

%\clearpage

\clearpage

\begin{appendices}

%%%Appendix A
\section{Proof of Theorem \ref{Thm:Price and User Association Equilibrium}} \label{Appen:Price and User Association Equilibrium}
We only discuss the case of $B_S, B_M > 0$. The sub-cases of either or both the variables being $0$ follow a similar logic with the obvious restriction of no users being served in the bands with no bandwidth.

1. We first show that the market always clears, i.e., all users would be served and the total rate supply is equal to the total rate demand.

Since the WiFi network on unlicensed spectrum is free to use, it's obvious that all fixed users would be served. If there are some mobile users that are not served yet, the SP can increase the price in macro-cells so that users in macro-cells would request less rate, leading to the SP having some redundant rate to serve more mobile users. The SP can thus use up its rate in macro-cells at a higher price since the unserved mobile users would fill in, which then leads to larger revenue.

On the other hand, if all users are served but there is still some redundant rate available in macro-cells or small-cells, the SP can decrease the price in corresponding cells so that users now request a higher rate. Since $ru^{'}(r)$, which is the revenue gained per user, increases with $r$, it's easy to see the SP can gain more revenue by doing so.

As a result of the market clearing property we have the following important conclusion. In both the macro-cells and the small-cells, the per-user rate equals the allocated total rate (bandwidth times the spectral efficiency) divided by the mass of customers associated with macro-cells and the same-cells. The price for access is given precisely by the inverse of the demand function $D(\cdot)$ at this per-user rate.

2. We then prove the price choice and user association equilibrium in the two different scenarios, given the fixed bandwidth allocation.

Assume macro-cells only serve mobile users. This then implies that $R_S \geq R_M$ so that the boundary point would correspond to the point at which $R_S=R_M$ holds. It can be determined using the following steps.
\begin{align}
&K_M=N_m\\
&u\Big( \frac{\lambda_SB_SR_0}{K_S}   \Big)  -  \frac{\lambda_SB_SR_0}{K_S}u^{'}\Big( \frac{\lambda_SB_SR_0}{K_S} \Big)
=u\Big( \frac{C_U}{N_f-K_S}   \Big)\\
&R_S=\frac{\lambda_SB_SR_0}{K_S}=R_M=\frac{B_MR_0}{K_M}
\end{align}
\noindent The above equation simplify to
\begin{equation}
B_S=\frac{\kappa N_fB_MR_0-N_mC_U}{\kappa\lambda_SN_mR_0} =: B_S^0.
\end{equation}

Therefore if $B_S$ is larger than $B_S^0$, it's easy to see that the user equilibrium must be such that macro-cells only serve mobile users. In contrast, if $B_S$ is smaller than $B_S^0$, then $R_S<R_M$ holds assuming macro-cells only serve mobile users. As a result,  some fixed users would have the incentive to associate with macro-cells and we would next prove that in this case at user equilibrium it's indeed the case that some fixed users would associate with macro-cells.

%{\color{red} 3. Finally we prove that for mixed service equilibrium, prices in macro-cells and small-cells must be equal.

Suppose $B_S<B_S^0$ and if macro-cells only serve mobile users such that the market clears, then we have $p_M<p_S$. The SP can then increases the price to $p_M^{'}$, where $p_M<p_M^{'}<p_S$, so that the mobile obtain a smaller rate creating some spare capacity. As a result, some fixed users in small-cells and WiFi network would switch to macro-cells, denote the total mass as $\delta$ and the mass of fixed users from small-cells switching being $\delta^\prime \leq \delta$. Note that before the price change $K_M=N_m$. The resulting revenue of the SP would then be :
\begin{align*}
S=&
B_MR_0u^{'}\Big(R_M \frac{N_m}{N_m+\delta}\Big)\\
&\qquad + \lambda_SB_SR_0
u^{'}\Big(R_S \frac{K_S}{K_S-\delta^\prime}\Big)
\end{align*}
\noindent By Lemma \ref{Lemma:Fixed user equilbrium} we can rewrite the revenue as:
\begin{align*}
S=&
B_MR_0u^{'}\Big(R_M \frac{N_m}{N_m+\delta}\Big)\\
&\qquad + \lambda_SB_SR_0
u^{'}\Big(R_S \frac{N_f}{N_f-\delta}\Big)
\end{align*}
\noindent Then we have:
\begin{align*}
\frac{\partial S}{\partial \delta}&=
-{R_M^{'}}^2u^{''}(R_M^{'})+\frac{\lambda_SB_SR_0}{N_f-\delta}
R_S^{'}u^{''}(R_S^{'})\\
& = -{R_M^{'}}^2u^{''}(R_M^{'})+ \frac{K_S}{N_f} \frac{\lambda_SB_SR_0}{K_S} \frac{N_f}{N_f-\delta} R_S^{'}u^{''}(R_S^{'})\\
&>-{R_M^{'}}^2u^{''}(R_M^{'})+{R_S^{'}}^2u^{''}(R_S^{'}),
\end{align*}
where $R_S^\prime= R_S\tfrac{K_S}{K_S-\delta^\prime}=R_S \tfrac{N_f}{N_f-\delta}$ and $R_M^\prime = R_M \tfrac{N_m}{N_m+\delta}$ are the new per user rates in the small-cells and macro-cells, respectively, after the shift of $\delta$ mass of fixed users to macro-cells.

Based on our assumptions, $r^2u^{''}(r)=-\alpha r^{1-\alpha}$ decreases with $r$, therefore as long as $R_S^{'}<R_M^{'}$, i.e., $p_M^{'}<p_S^{'}$, $S$ always increases with $\delta$. As a result, it's always better for macro-cells to serve some fixed users in this case and the optimal price choice is $p_M=p_S$.

In mixed service scenario, we therefore have the following equations:
\begin{align}
\begin{split}
u\Big(\frac{C_S}{K_S}\Big)-p_S\frac{C_S}{K_S}& =u\Big(\frac{C_U}{K_U}\Big) =u\Big(\frac{C_M}{K_M}\Big) - p_M\frac{C_M}{K_M},\\
K_U+K_S+K_M& =N_m+N_f,\\
D(p_M)=R_M&=D(p_S)=R_S=\frac{C_S+C_M}{K_M+K_S}.
\label{eq:mixed}
\end{split}
\end{align}

Using Lemma \ref{Lemma:Fixed user equilbrium} we can get:
%The corresponding user association equilibrium can therefore be calculated as follows:\\
%\noindent 1. Mixed service scenario:
\begin{align}\label{Appen:Eqn:Mixed Service}
\begin{split}
K_U &=(N_f+N_m)\frac{C_U}{C_U+\kappa(C_M+C_S)},\\
K_M &=(N_f+N_m)\frac{\kappa C_M}{C_U+\kappa(C_M+C_S)},\\
K_S &=(N_f+N_m)\frac{\kappa C_S}{C_U+\kappa(C_M+C_S)}.
\end{split}
\end{align}
which gives the number of active users in terms of the network capacities.

Similarly, for the separate service scenario,
writing the analogous conditions to (\ref{eq:mixed}) we can get:
\begin{align}\label{Appen:Eqn:Separate Service}
\begin{split}
K_M&=N_m, \quad K_S=N_f\frac{\kappa C_S}{\kappa C_S+C_U},\\
K_U&=N_f\frac{C_U}{\kappa C_S+C_U}.
\end{split}
\end{align}

%%%Appendix B
\section{Proof of Theorem \ref{Thm:Optimal Bandwidth Allocation}} \label{Appen:Optimal Bandwidth Allocation}
1. We first prove that the optimal bandwidth allocation cannot occur at mixed service scenario. The revenue of the SP under the mixed service scenario is:
\begin{align}
S=&(B_M+\lambda_SB_S)R_0u^{'}\Big(\frac{(B_M+\lambda_SB_S)R_0}{K_M+K_S}\Big)\\
=&(B_M+\lambda_SB_S)R_0u^{'}\Big(\frac{C_U+\kappa(B_M+\lambda_SB_S)R_0}{\kappa(N_m+N_f)}\Big)\\
=&(N_m+N_f)Ru^{'}(R)-\frac{C_U}{\kappa}u^{'}(R)
\end{align}
where $R:=\frac{C_U+\kappa(B_M+\lambda_S B_S)R_0}{\kappa(N_m+N_f)}$ is the average rate in both macro-cells and small-cells.

Based on our assumptions, $Ru^{'}(R)$ increases with $R$ and $u^{'}(R)$ decreases with $R$, therefore it's always beneficial to allocate more bandwidth to small-cells: since $\lambda_S > 1$, $R$ increases with $B_S$. This means the optimal point cannot exist at a mixed service scenario.

2. We then prove the optimal bandwidth allocation scheme in separate service scenario. The revenue of the SP at separate service equilibrium is:
\begin{align}
S=&B_MR_0u^{'}\Big(\frac{B_MR_0}{K_M}\Big) +\lambda_SB_SR_0u^{'}       \Big(\frac{\lambda_SB_SR_0}{K_S}\Big)\\
=&N_mR_Mu^{'}(R_M)+N_fR_Su^{'}(R_S)-\frac{C_U}{\kappa}u^{'}(R_S)
\end{align}
where $R_M=\frac{B_MR_0}{N_m}, R_S=\frac{\lambda_SB_SR_0}{K_S}=\frac{\kappa\lambda_SB_SR_0+C_U}{\kappa N_f}$ are the average rate in macro-cells and small-cells, respectively.

It's easy to verify for $\alpha$-fair utility functions, $R_Mu^{'}(R_M), R_Su^{'}(R_S)$ are concave increasing functions with respect to $B_M$ and $B_S$, respectively. Furthermore, $-u^{'}(R_S)$ is also a concave increasing function with respect to $B_S$. As a result, $S$ is a increasing with either $B_S$ or $B_M$. Therefore at optimal point the SP uses up all its total bandwidth and we have $B_S+B_M=1$. This further means $S$ is concave with $B_S$. As a result, the optimal point occurs at the point which uses up the total bandwidth and equalizes the marginal revenue increase with respect to per unit bandwidth increase in both macro-cells and small-cells, which can be achieved by straightforward calculation given below:

\begin{align}
\begin{split}
& \frac{1-\alpha}{\lambda_S}\big( \frac{B_M^\text{rev}R_0}{N_m}    \big)^{-\alpha}  =  \Bigg[ \big( \frac{\kappa\lambda_SB_S^\text{rev}R_0+C_U}{\kappa N_f}   \big)^{-\alpha} \\
& \qquad \quad -\alpha\big(\frac{\kappa\lambda_SB_S^\text{rev}R_0+C_U}{\kappa N_f}\big) ^{-\alpha} \frac{\kappa\lambda_SB_S^\text{rev}R_0}{\kappa\lambda_SB_S^\text{rev}R_0+C_U}         \Bigg] ,  \\
& \qquad B_S^{\text{rev}}+B_M^{\text{rev}}  =B, \qquad B_S^\text{rev}, B_M^\text{rev} \geq 0.
\end{split}
\tag{P1} \label{Eqn:Revenue Opt with unlicensed}
\end{align}

%%%Appendix C
\section{Proof of Theorem \ref{Thm:Comparison}} \label{Appen:Comparison}
We first make the following definitions:
\begin{align}
A_1=&\Big[
-\frac{\alpha \kappa\lambda_SB_S}{\kappa\lambda_SB_S+\lambda_UB_U} \bigg(\frac{\kappa\lambda_SB_S+\lambda_UB_U}{\kappa N_f}R_0\bigg) ^{-\alpha}\\
&  \qquad   +\bigg( \frac{\kappa\lambda_SB_S+\lambda_UB_U}{\kappa N_f} R_0  \bigg)^{-\alpha}     \Big] \lambda_S\\
A_2=&(1-\alpha)\lambda_S\bigg( \frac{\lambda_SB_SR_0}{N_f}\bigg) ^{-\alpha}
\end{align}

We only need to compare $A_1$ and $A_2$ at $\tilde{B}_S^\text{rev}$. By explicit calculation, this is given by:
\begin{equation}
A_1-A_2=\frac{M}{(\kappa\lambda_S\tilde{B}_S^\text{rev}R_0+C_U)^{\alpha+1}(\lambda_S\tilde{B}_S^\text{rev}R_0)^{\alpha}}
\end{equation}
where $M=(1-\alpha)\big[(\kappa\lambda_S\tilde{B}_S^\text{rev}R_0)^{\alpha+1}- (\kappa\lambda_S\tilde{B}_S^\text{rev}R_0+C_U)^{\alpha+1}\big]-C_U(\kappa\lambda_S\tilde{B}_S^\text{rev}R_0)^{\alpha}$.

It's easy to verify that $M$ first increases with $C_U$ and then decreases with $C_U$. Moreover, $M=0$ when $C_W=0$. Therefore we only need to determine the other zero-crossing point $C_U^{th}$ by letting $M=0$. The conclusions in Theorem \ref{Thm:Comparison} then follow in a straightforward manner.

%%%Appendix D
\section{Proof of Theorem \ref{Thm:Social Welfare Maximization}} \label{Appen:Social Welfare Maximization}
1. We first show that the market always clears, i.e., all users would be served and the total rate supply is equal to the total rate demand.

Since the WiFi network is free to use, again it's obvious that all fixed users would be served. If there are some mobile users that are not served yet, the SP can increase the price in macro-cells so that users in macro-cells would request less rate and therefore it has some redundant rate to serve the mobile users. Since the function $K_Mu(\frac{C_M}{K_M})$ increases with $K_M$, the social welfare is thus increased.

On the other hand, if all users are served but there is still some redundant rate available in macro-cells or small-cells. The SP can decrease the price in corresponding cells so that users now request higher rate, which naturally leads to higher social welfare.

2. We then prove the price choice and users association equilibrium in two different scenarios, depending on the fixed bandwidth allocation.

Similar to the analysis for revenue maximization, we first prove that at certain fixed bandwidth allocation, if macro-cells only serve mobile users and mobile users achieve larger rate than fixed users in small-cells, then the final user association should be that some fixed users associate with macro-cells and the price in macro-cells and small-cells should be the same.

Suppose macro-cells only serve mobile users and we have  $p_M<p_S, R_M>R_S$. The SP can increase the macro-cell price to $p_M^{'}$, where $p_M<p_M^{'}<p_S$. As a result, some fixed users in small-cells and WiFi network would switch to macro-cells, denote the mass of these fixed users as $\delta$ and $N_t=N_m+N_f=K_M+K_S+K_U$. By Lemma \ref{Lemma:Fixed user equilbrium} we can show that the social welfare would then be :
\begin{align}
\begin{split}
\text{SW}=&(K_M+\delta)u\Big(  R_M\frac{K_M}{K_M+\delta}  \Big) \\
&\quad + K_S\frac{N_t-K_M-\delta}{N_t-K_M} u\Big( R_S\frac{N_t-K_M}{N_t-K_M-\delta} \Big)\\
&\quad +K_U\frac{N_t-K_M-\delta}{N_t-K_M} u\Big( R_U\frac{N_t-K_M}{N_t-K_M-\delta} \Big).
\end{split}
\end{align}
\noindent Then we have:
\begin{align}
& \notag \frac{\partial \text{SW}}{\partial \delta}=
u(R_M^{'})-R_M^{'}u^{'}(R_M^{'})-\frac{K_S}{N_t-K_M}\Big[ u(R_S^{'})-\\
&\qquad R_S^{'}u^{'}(R_S^{'})\Big]-
\frac{K_U}{N_t-K_M}\Big[ u(R_U^{'})-R_U^{'}u^{'}(R_U^{'})\Big]
\end{align}

For the $\alpha$-fair utility functions we use, it's easy to verify that $u(r)-ru^{'}(r)$ increases with $r$. We also have $R_M^{'}>R_S^{'}>R_U^{'}$, thus
\begin{align*}
\notag \frac{\partial \text{SW}}{\partial \delta} &>
u(R_M^{'})-R_M^{'}u^{'}(R_M^{'})- \big[ u(R_S^{'})-R_S^{'}u^{'}(R_S^{'})\big]\\
&>0.
\end{align*}

Therefore the social welfare increases with $\delta$ in mixed service scenario as long as $p_M<p_S$. As a result, in this case some fixed users would associate with macro-cells, and the optimal price choice will be $p_M=p_S$.

3. Next, we show that the optimal bandwidth allocation is unique and only occurs at separate service scenario. The social welfare at mixed service scenario is:
\begin{equation}
\text{SW}= (K_M+K_S)u(R)+K_Uu(R_U)
\end{equation}
where $R=\frac{[\lambda_UB_U+\kappa(B_M+\lambda_SB_S)]R_0)}{\kappa(N_m+N_f)}$, $R_U= \frac{[\lambda_UB_U+\kappa(B_M+\lambda_SB_S)]R_0)}{N_m+N_f}$ are the average user rate in licensed and unlicensed spectrum, respectively. Therefore we have:
\begin{align}
\notag \text{SW}=& (K_M+K_S)u(R)+K_Uu(\kappa R)\\
\notag =&(N_m+N_f)u( R)-K_U(u(R)-u(\kappa R))\\
=&(N_m+N_f)u( R)-C_U\frac{u(R)-u(\kappa R)}{\kappa R}
\end{align}

It's easy to see that the first term is increasing with $R$. For $\alpha$-fair utility functions we use, it's easy to verify the second term is decreasing with $R$. Therefore the social welfare increases with $R$, which leads to the conclusion that it's always better to invest more bandwidth to small-cells. As a result, the optimal point cannot occur at mixed service scenario.

We then prove that the optimal bandwidth allocation is unique at separate service scenario.
\begin{align}
\notag &\text{SW}=N_mu(R_M)+K_Su(R_S)+K_Uu(R_U)\\
\notag & R_S=\frac{\kappa\lambda_SB_SR_0+C_U}{\kappa N_f}, R_U=\kappa R_S\\
& K_S=\frac{\lambda_SB_SR_0}{R_S}, K_U=\frac{C_U}{R_U}
\end{align}

We then have:
\begin{equation}
\text{SW}= N_mu(R_M)+ N_fu(R_S)+\frac{C_U}{\kappa}[\frac{u(\kappa R_S)-u(R_S)}{R_S}]
\end{equation}

Since $N_mu(R_M)$ is concave increasing with $R_M$ and both $N_fu(R_S)$, $\frac{u(\kappa R_S)-u(R_S)}{R_S}= -R_S^{-\alpha}$ are concave increasing with $R_S$. It's easy to verify that social welfare is increasing with either $B_S$ or $B_M$ and therefore at optimal point we have $B_S+B_M=B$. By this we can further show that it is also a concave function with $B_S$ and therefore the optimal point is unique. At this optimal point, SP uses up the total bandwidth and equalizes the marginal social welfare increase of mobile users and fixed users with respect to per unit of bandwidth increase in macro-cells and small-cells. The optimal bandwidth allocation can be calculated as follows:

\begin{displaymath}\tag{P3} \label{Eqn:SW Opt with unlicensed}
\left\{
\begin{array}{ll}
\frac{ (N_f)^{\alpha}\lambda_S\big((\kappa^{\alpha}+\kappa) C_U+ \kappa^{\alpha+1}\lambda_SB_S^{\text{sw}}R_0  \big)        }{ (\kappa\lambda_SB_S^{\text{sw}}R_0+C_U)^{\alpha+1}             } = (\frac{ B_M^{\text{sw}}R_0}{N_m  })^{-\alpha}, \\
B_S^{\text{sw}}+B_M^{\text{sw}}=1, \qquad B_S^{\text{sw}}, B_M^{\text{sw}} \geq 0.
\end{array}
\right.
\end{displaymath}

4. Last, we prove $B_S^{\text{sw}}<\tilde{B}_S^{\text{sw}}$. We first define some notations.
\begin{align}
\notag &A_3=\frac{(N_f)^{\alpha}\lambda_S\big[(\kappa^{\alpha}+\kappa) C_U+ \kappa^{\alpha+1}\lambda_S\tilde{B}_S^{*}R_0\big]}
{(\kappa\lambda_S\tilde{B}_S^{*}R_0+C_U)^{\alpha+1}} \\
&A_4=\lambda_S\big( \frac{\lambda_S\tilde{B}_S^{**}R_0}{N_f}\big) ^{-\alpha}
\end{align}
%{\color{blue} You use $\tilde{B}_S^*$ and $\tilde{B}_S^{**}$.}

We only need to compare $A_3$ and $A_4$ at $\tilde{B}_S^{\text{sw}}$. It turns out that:
\begin{align}
\notag &\frac{A_3-A_4}{\lambda_SR_0(N_f)^{\alpha}}=\frac{
M}
{(\lambda_S\tilde{B}_S^{\text{sw}}R_0)^{\alpha}(\kappa\lambda_S\tilde{B}_S^{\text{sw}}R_0+C_U)^{\alpha+1}}\\
\notag &M=(\kappa^{\alpha}+\kappa)C_U(\lambda_S\tilde{B}_S^{\text{sw}}R_0)^{\alpha}+\kappa^{\alpha+1}
(\lambda_S\tilde{B}_S^{\text{sw}}R_0)^{\alpha+1}\\
&-(\kappa\lambda_S\tilde{B}_S^{\text{sw}}R_0+C_U)^{\alpha+1}
\end{align}
%{\color{blue} Now it is just $B_S$.}

We then have:
\begin{align}
\notag \frac{\partial M}{\partial C_U}=&
(\alpha+1)\Big[  (\kappa\lambda_S\tilde{B}_S^{\text{sw}}R_0)^{\alpha} - (\kappa\lambda_S\tilde{B}_S^{\text{sw}}R_0+C_U)^{\alpha}   \Big]\\
<& 0, \quad \forall C_U>0.
\end{align}

Since when $C_U=0$, $A_3=A_4$. Therefore $A_3<A_4$ when $C_U>0$. As a result, compared with the scenario without unlicensed spectrum, the SP should always invest less bandwidth to small-cells in terms of social welfare maximization.

%%%Appendix E
\section{Proof of Theorem \ref{Thm:Price NE}} \label{Appen:Price NE}
The proof is essentially the same as the proof for the price choice and user association equilibrium for monopoly service provider scenario. The only difference is that we need to prove the prices must be equal across all small-cells or macro-cells with multiple SPs.

Suppose one SP $i$ has lower small-cell price $p_{i, S}$ than the other SP $j$. SP $i$ can increase the price to $p_{i, S}^{'}$ satisfying $p_{i, S}<p_{i, S}^{'}<p_{j, S}$. As a result, SP $i$ would attract some users which used to associate with SP $j$'s small-cell or some users from unlicensed access and still use up all the rate with a higher price. Therefore SP $i$ can increase its revenue by doing so. Thus, at price equilibrium all small-cell price must be equal. The proof for all macro-cell prices must be equal is done similarly.
%{\color{blue} For fixed users you need to be careful about users moving from unlicensed spectrum too. That's missing in your discussion above.}

%%%Appendix F
\section{Proof of Theorem \ref{Thm:Existence and Uniqueness of NE}} \label{Appen:Existence and Uniqueness of NE}
We prove the theorem in the following steps.

1. First, we prove that no Nash equilibrium exists at mixed service scenario. It's easy to see the revenue of SP $i$ in this case is:
\begin{align}
&S_i=(B_{i,M}+\lambda_SB_{i,S})R_0u^{'}\Big[  \frac{\sum\limits_{j=1}^{N}(B_{j,M}+\lambda_SB_{j,S})R_0}{K_M+K_S}   \Big]\\
&=(K_M+K_S)Ru^{'}(R)-\sum\limits_{j\ne i}^{N}(B_{j,M}+\lambda_SB_{j,S})R_0u^{'}(R)\\
&K_M+K_S=N_m+N_f-
\frac{(N_m+N_f)C_U}{C_U+\kappa\sum\limits_{j=1}^{N}(B_{j,M}+\lambda_SB_{j,S})R_0}\\
&R=\frac{C_U+\kappa\sum\limits_{j=1}^{N}(B_{j,M}+\lambda_SB_{j,S})R_0}{\kappa(N_m+N_f)}
\end{align}

It's easy to see that $R$ and $K_M+K_S$ both increase with $B_{i,S}$. Meanwhile, $Ru^{'}(R)$ increases with $R$ and $u^{'}(R)$ decreases with $R$. As a result, $S_i$ increases with $B_{i,S}$ and every SP would have incentive to invest more bandwidth to small-cells, which would finally push the equilibrium to separate service scenario. Therefore no Nash equilibrium exists in mixed service scenario.

2. We next prove that there always exists a Nash equilibrium at separate service scenario. The revenue of SP $i$ in this case is:
\begin{align}
\notag S_i=&S_{i,M}+S_{i,S}\\
\notag =&B_{i,M}R_0u^{'}(R_M)+\lambda_SB_{i,S}R_0u^{'}(R_S)\\
=&B_{i,M}R_0u^{'}\Big(   \frac{\sum\limits_{j=1}^{N}B_{j,M}R_0}{N_m}    \Big) +
\lambda_SB_{i,S}R_0u^{'}\Big(   \frac{\sum\limits_{j=1}^{N}\lambda_SB_{j,S}R_0}{K_S}    \Big)
\end{align}

It's easy to verify:
\begin{align}
\notag \frac{\partial S_{i,M}}{\partial B_{i,M}}=&R_0\big[ u^{'}(R_M)+\frac{B_{i,M}R_0}{N_m}u^{''}(R_M)  \big]\\
\notag =&R_0\big[ u^{'}(R_M)+R_Mu^{''}(R_M)-\frac{\sum\limits_{j\ne i}B_{j,M}R_0}{N_m}u^{''}(R_M)  \big]\\
>&0
\end{align}
\begin{align}
\notag \frac{\partial S_{i,S}}{\partial B_{i,S}}=&\lambda_SR_0\big[ u^{'}(R_S)+\frac{\lambda_SB_{i,S}R_0}{N_f}u^{''}(R_S)  \big]\\
\notag =&\lambda_SR_0\big[ u^{'}(R_S)+\frac{\sum\limits_{j=1}^{N}\lambda_SB_{j,S}R_0}{N_f}u^{''}(R_S)\\
\notag &-\frac{\sum\limits_{j\ne i}\lambda_SB_{j,S}R_0}{N_f}u^{''}(R_S)  \big]\\
\notag =&\lambda_SR_0\big[ u^{'}(R_S)+R_Su^{''}(R_S)\\
&-\frac{\lambda_SC_U}{\kappa N_f}u^{''}(R_S)-\frac{\sum\limits_{j\ne i}\lambda_SB_{j,S}R_0}{N_f}u^{''}(R_S)  \big]>0
\end{align}

Since $u^{'}(r)+ru^{''}(r)$ decreases with $r$ and $u^{''}(r)$ increases with $r$, it's easy to verify $S_{i,M}$ and $S_{i,S}$ are both concave increasing with $B_{i,M}$ and $B_{i,S}$ respectively. This also indicates all SPs would always use up their total bandwidth, i.e., $B_{i,S}+B_{i,M}=B_I$. By this we can further show $S_i$ is a concave function with $B_{i,S}$.  Moreover, the constraint on separate service scenario are linear with $B_{i,S}$. We can then apply Rosen's theorem on concave games \cite{Rosen65} to prove the existence of Nash equilibrium.

3. For the third step, we prove that fixed users in small-cells achieve higher average rate than mobile users in macro-cells. This is equivalent to say $R_S>R_M$ at equilibrium.

We only need to rule out the possibility that $R_S=R_M$ since we already showed that the Nash equilibrium falls into separate service scenario. Denote the group of SPs that only allocate bandwidth to small-cells (macro-cells) as $G_S (G_M)$ and the group of SPs that allocate bandwidth to both cells as $G_{MS}$, then we have:
\begin{align}
\notag &\forall i\in G_S, B_{i,S}=B_i, B_{i,M}=0\\
&\forall j\in G_M \cup G_{MS}, \frac{\partial S_j}{B_{j,S}} \le 0
\end{align}

If $R_S=R_M=R$ holds, then $\forall j\in G_M \cup G_{MS}$, we have:
\begin{align}
\notag \frac{\partial S_j}{\partial B_{j,S}}=& \lambda_SR_0
\Big[ u^{'}(R_S)+\frac{\lambda_SB_{j,S}R_0}{N_f}u^{''}(R_S)\Big]\\
\notag &-R_0\Big[ u^{'}(R_M)+\frac{B_{j,M}R_0}{N_m}u{''}(R_M)  \Big] \\
\notag =& (\lambda_S-1)u^{'}(R)+\Big[\frac{\lambda_SB_{j,S}R_0}{N_f}
-\frac{B_{j,M}R_0}{N_m} \Big]u^{''}(R)\\
\le & 0
\end{align}

Therefore $\forall j\in G_M \cup G_{MS}$, we have:
\begin{equation}
\frac{\lambda_SB_{j,S}R_0}{N_f} \ge \frac{B_{j,M}R_0}{N_m}
\end{equation}

As a result, the following holds:
\begin{align}
R_S=&\sum\limits_{j\in G_M \cup G_{MS}}\frac{\lambda_SB_{j,S}R_0}{N_f}+ \sum\limits_{i\in G_S }\frac{\lambda_SB_iR_0}{N_f}>\\
&\sum\limits_{j\in G_M \cup G_{MS}}\frac{B_{j,M}R_0}{N_m} = R_M
\end{align}
Therefore we have a contradiction.

4. We then show that at Nash equilibrium it is impossible that one SP only allocates bandwidth to macro-cells while the other SP only allocates bandwidth to small-cells.

Suppose SP $i$ only allocates bandwidth to macro-cells while SP $j$ only allocates bandwidth to small-cells. Then we have:
\begin{align}
\notag \frac{\partial S_i}{\partial B_{i,S}} =& \lambda_SR_0
 u^{'}(R_S)-R_0\Big[ u^{'}(R_M)+\frac{B_{i,M}R_0}{N_m}u{''}(R_M)  \Big]\\ \le &0\\
\notag \frac{\partial S_j}{\partial B_{j,S}} =& \lambda_SR_0
\Big[ u^{'}(R_S)+\frac{\lambda_SB_{j,S}R_0}{N_f}u^{''}(R_S)\Big]
 -R_0 u^{'}(R_M) \\
 \ge & 0
\end{align}
which would yield:
\begin{align}
-\frac{B_{i,M}R_0}{N_m}u{''}(R_M) \le \lambda_S\frac{\lambda_SB_{j,S}R_0}{N_f}u^{''}(R_S)
\end{align}

Clearly we have a contradiction then.

5. After step 4, it's clear that there are only five possible Nash equilibrium types:
\begin{itemize}
\item[1)]
Small-only Nash equilibrium(SNE): All SPs only allocate bandwidth to small-cells.
\item[2)]
Macro-only Nash equilibrium(MNE): All SPs only allocate bandwidth to macro-cells.
\item[3)]
Macro-Small Nash Equilibrium (MSNE): All SPs allocate bandwidth to both macro- and small-cells.
\item[4)]
Macro-Preferred Nash Equilibrium (MPNE): Some SPs allocate bandwidth to both small-- and macro-cells while the other SPs only allocate bandwidth to macro-cells.
\item[5)]
Small-Preferred Nash Equilibrium (SPNE): Some SPs allocate bandwidth to both small-- and macro-cells while the other SPs only allocate bandwidth to small-cells.
\end{itemize}

We next prove that SNE cannot exist.

The marginal revenue increase with respect to per unit bandwidth increase in macro-cells for SP $i$ is given by:
\begin{equation}
\frac{\partial S_i}{\partial B_{i,M}} = R_0
\Big[ u^{'}(R_M)+\frac{B_{i,M}R_0}{N_m}u^{''}(R_M)\Big]
\end{equation}

It can be easily shown that for $\alpha$-fair utility functions, the marginal revenue increase with respect to per unit bandwidth increase in macro-cells goes to infinity when $B_M$ is near 0. As a result, SNE cannot exist.

However, the above argument doesn't apply to MNE. At first glance, it seems that when all SPs only allocate bandwidth to macro-cells, the marginal revenue increase with respect to per unit bandwidth in small-cells also goes to infinity when $B_S$ is near 0. Actually, the marginal revenue increase does go to infinity when $R_S$ is near 0. However, $R_S$ doesn't go to 0 when $B_S$ goes to 0. In fact, $R_S$ is discontinuous at the point 0. We have the following:
\begin{align}
&R_S=\frac{\kappa\lambda_SB_SR_0+C_U}{\kappa N_f},& B_S>0;\\
&R_S=0,  &B_S=0.
\end{align}

Therefore as $B_S\rightarrow 0^+, R_S\rightarrow \frac{C_U}{\kappa N_f} >0$.

6. The fact that SPNE cannot exist requires more work and in this part we would focus on it. At SPNE, we know that there exist SPs $i,j$ such that:
\begin{align}
& \frac{\partial S_i}{\partial B_{i,S}}\ge 0,  B_{i,S}=B_i, B_{i,M}=0\\
& \frac{\partial S_j}{\partial B_{j,S}} = 0, B_{j,S}>0, B_{j,M}>0
\end{align}

We therefore have:
\begin{align}
& \lambda_F\Big[u^{'}(R_S)+ \frac{\lambda_SB_{i,S}R_0}{N_f}u{''}(R_S)  \Big] \ge
u^{'}(R_M)\\
\notag & \lambda_S\Big[u^{'}(R_S)+ \frac{\lambda_SB_{i,S}R_0}{N_f}u{''}(R_S)  \Big] =
u^{'}(R_M)+\\
&\frac{B_{i,M}R_0}{N_m}u^{''}(R_M)
\end{align}

From the first inequality above, we can easily conclude that:
\begin{equation} \label{Eqn:first inequality}
\lambda_Su^{'}(R_S)>u^{'}(R_M)
\end{equation}

Next, we consider the group of service providers that allocate bandwidth to both cells, denoted as $G_{MS}$ and assume $|G_{MS}|=L$.
\begin{equation}
\forall j\in G_{MS}, \frac{\partial S_j}{\partial B_{j,S}}=0, B_{j,S}>0, B_{j,M}>0
\end{equation}

We then have:
\begin{align}
\notag &\lambda_S\Big[Lu^{'}(R_S)+ \frac{\lambda_S\sum\limits_{j\in G_{MS}}B_{j,S}R_0}{N_f}u{''}(R_S)  \Big] =
\\
&Lu^{'}(R_M)+\frac{\sum\limits_{j\in G_{MS}}B_{j,M}R_0}{N_m}u^{''}(R_M)
\end{align}

It's easy to get:
\begin{align}
\notag &\lambda_S\Big[Lu^{'}(R_S)+ \frac{\lambda_S\sum\limits_{j\in \mathcal{N}}B_{j,S}R_0}{N_f}u{''}(R_S)  \Big] <
\\
&Lu^{'}(R_M)+R_Mu^{''}(R_M)
\end{align}

Together with inequality \ref{Eqn:first inequality}, we can get the second inequality:
\begin{align}
\notag &\lambda_S\Big[u^{'}(R_S)+ \frac{\lambda_S\sum\limits_{j\in \mathcal{N}}B_{j,S}R_0}{N_f}u{''}(R_S)  \Big] <\\
&u^{'}(R_M)+R_Mu^{''}(R_M)
\end{align}

For $\alpha$-fair utility functions, we have:
\begin{equation}
u^{'}(R_M)+R_Mu^{''}(R_M)=(1-\alpha)u^{'}(R_M)
\end{equation}

However, we also have:
\begin{align}
\notag &\lambda_S\Big[u^{'}(R_S)+ \frac{\lambda_S\sum\limits_{j\in \mathcal{N}}B_{j,S}R_0}{N_f}u{''}(R_S) \Big]-(1-\alpha)\lambda_Su^{'}(R_S)\\
\notag &=\lambda_S\Big[\alpha u^{'}(R_S)+ \frac{\lambda_S\sum\limits_{j\in \mathcal{N}}B_{j,S}R_0}{N_f}u{''}(R_S)\Big]\\
\notag &>\lambda_S\Big[\alpha u^{'}(R_S)+ R_Su{''}(R_S)\Big]\\
&=\lambda_S\alpha R_S^{-\alpha}-\lambda_S\alpha R_S^{-\alpha}=0
\end{align}

Therefore the two inequalities have a contradiction. As a result, SPNE cannot exist.

7. We next prove that MSNE, MNE and MPNE cannot coexist. First it's easy to verify that at MSNE, MNE and MPNE,  we have the following:
\begin{align}
\notag \text{MSNE}:\quad &\lambda_S\Big[   Nu^{'}(R_S)+\frac{\lambda_S\sum\limits_{i\in \mathcal{N}} B_{i,S}R_0}{N_f} u^{''}(R_S)    \Big]\\
& = Nu^{'}(R_M)+R_Mu^{''}(R_M)\\
\notag \text{MNE or MPNE}:\quad &\lambda_S\Big[   Nu^{'}(R_S)+\frac{\lambda_S\sum\limits_{i\in \mathcal{N}} B_{i,S}R_0}{N_f} u^{''}(R_S)    \Big]\\
& \le Nu^{'}(R_M)+R_Mu^{''}(R_M)
\end{align}

We can show that the LHS is a decreasing function with $R_S$, since we have:
\begin{equation}
R_S=\frac{C_U+\kappa\lambda_S\sum\limits_{i\in \mathcal{N}}B_{i,S}R_0 }{\kappa N_f}
\end{equation}

Therefore we have:
\begin{align}
\text{LHS}=&\lambda_S\big[   u^{'}(R_S)+R_Su^{''}(R_S)   \big] +(N-1)\lambda_Su^{'}(R_S)\\
&-\frac{\lambda_SC_U}{\kappa N_f}u^{''}(R_S)
\end{align}
which is decreasing with $R_S$.

Similarly, RHS is also a decreasing function with $R_M$.

Now suppose for the same set of parameters, we have one MNE or MPNE and another MSNE, denote the corresponding bandwidth allocation profile as $\mathbf{B}$ and $\bar{\mathbf{B}}$, respectively. Then we must have:
\begin{equation}
\bar{R_S} \le R_S, \bar{R_M}\ge R_M
\end{equation}

Now it is clearly shown that MNE and MSNE cannot coexist since for MNE we have $R_M>\bar{R_M}$.

If $\mathbf{B}$ corresponds to MPNE, we can conclude that at MSNE, some SPs must have less bandwidth allocation to small-cells than that of MPNE. Denote this group of SPs as $G_{S-}$ and assume $|G_{MS-}|=L$, we have:
\begin{equation}
\forall j\in G_{S-}, \frac{\partial S_j}{\partial B_{j,S}}=0, \frac{\partial \bar{S_j}}{\partial \bar{B}_{j,S}}\le 0
\end{equation}

Summing up, we get:
\begin{align}
\notag &\lambda_S\Big[   Lu^{'}(R_S)+\frac{\lambda_S\sum\limits_{j\in G_{S-}} B_{j,S}R_0}{N_f} u^{''}(R_S)    \Big]   =\\
&  Lu^{'}(R_M)+\frac{\sum\limits_{j\in G_{S-}} B_{j,M}R_0}{N_m}u^{''}(R_M)\\
\notag &\lambda_S\Big[   Lu^{'}(\bar{R_S})+\frac{\lambda_S\sum\limits_{j\in G_{S-}} \bar{B_{j,S}}R_0}{N_f} u^{''}(\bar{R_S})    \Big]   \le\\
&  Lu^{'}(\bar{R_M})+\frac{\sum\limits_{j\in G_{S-}} \bar{B_{j,M}}R_0}{N_m}u^{''}(\bar{R_M})
\end{align}

Rearranging some of the terms, we have:
\begin{align}
\notag &\lambda_S\Big[   Lu^{'}(R_S)+\frac{\lambda_S\sum\limits_{j\in \mathcal{N}} B_{j,S}R_0}{N_f} u^{''}(R_S)    \Big] = \\
\notag &  Lu^{'}(R_M)+ R_Mu^{''}(R_M)+  \lambda_S\frac{\lambda_S\sum\limits_{j\notin {G_{S-}}} B_{j,S}R_0}{N_f} u^{''}(R_S)\\
&-\frac{\sum\limits_{j\notin G_{S-}} B_{j,M}R_0}{N_m}u^{''}(R_M)\\
\notag &\lambda_S\Big[   Lu^{'}(\bar{R_S})+\frac{\lambda_S\sum\limits_{j\in \mathcal{N}} \bar{B_{j,S}}R_0}{N_f} u^{''}(\bar{R_S})    \Big] \le \\
\notag &  Lu^{'}(\bar{R_M})+ \bar{R_M}u^{''}(\bar{R_M})+  \lambda_S\frac{\lambda_S\sum\limits_{j\notin {G_{S-}}} \bar{B_{j,S}}R_0}{N_f} u^{''}(\bar{R_S})\\
&-\frac{\sum\limits_{j\notin G_{S-}} \bar{B_{j,M}}R_0}{N_m}u^{''}(\bar{R_M})
\end{align}

However, we also have:
\begin{equation}
\forall j\not\in G_{S-}, \bar{B_{j,S}}\ge B_{j,S}, \bar{B_{j,M}}\le B_{j,M}
\end{equation}

We already showed that LHS decreases with $R_S$ and the first two terms on RHS decreases with $R_M$. For $\alpha$-fair utility functions, $u^{''}(r)<0$ increases with $r$. Combining with the fact that $\bar{R_S} \le R_S, \bar{R_M}\ge R_M$ and noticing that at least one of the inequalities must be strict inequality, we can also conclude:
\begin{align}
\notag &\lambda_S\Big[   Lu^{'}(\bar{R_S})+\frac{\lambda_S\sum\limits_{j\in \mathcal{N}} \bar{B_{j,S}}R_0}{N_f} u^{''}(\bar{R_S})    \Big] > \\
\notag &  Lu^{'}(\bar{R_M})+ \bar{R_M}u^{''}(\bar{R_M})+  \lambda_S\frac{\lambda_S\sum\limits_{j\notin {G_{S-}}} \bar{B_{j,S}}R_0}{N_f} u^{''}(\bar{R_S})\\
&-\frac{\sum\limits_{j\notin G_{S-}} \bar{B_{j,M}}R_0}{N_m}u^{''}(\bar{R_M})
\end{align}

%{\color{blue} Sometime you use $G_{S-}$ and other times $\mathcal{G}_{S-}$.}

Clearly we have a contradiction then. As a result, MSNE and MPNE cannot coexist.

We then only need to show MNE and MPNE cannot coexist. It can be proved in a similar way as we proved MSNE and MPNE cannot coexist introduced above. We now focus on the group of SPs which decrease bandwidth allocation to small-cells and apply the same procedures to get a contradiction.

8. Finally, we need to show that within MSNE, MNE or MPNE, the Nash equilibrium is unique.

The uniqueness of MNE is trivial.

The uniqueness of MPNE can be proved in a similar way in which we proved MSNE and MPNE cannot coexist. Here we don't repeat the steps.

The uniqueness of MSNE can also be proved similarly. However, here we use another method. It's easy to see that at MSNE, we have the following system of equations.
\begin{displaymath}
\left\{
\begin{array}{ll}
\lambda_S[   Nu^{'}(R_S)+\frac{\lambda_S\sum\limits_{i\in \mathcal{N}} B_{i,S}R_0}{N_f} u^{''}(R_S)    ]\\
 = Nu^{'}(R_M)+R_Mu^{''}(R_M)\\
\frac{\lambda_S^2\Delta B_{ij,S}}{N_f}u^{''}(R_S)=\frac{\Delta B_{ij,M}}{N_m}u^{''}(R_M)
\end{array}
\right.
\end{displaymath}
where $\Delta B_{ij,S}=B_{i,S}-B_{j,S}$ is the difference of bandwidth allocation to small-cells between SP $i$ and SP $j$, the same
for $\Delta B_{ij,M}$.

By the monotonicity of both LHS and RHS with respect to $R_S$ and $R_M$, the first equation we can uniquely determine $\sum\limits_{i=1}^{N}B_{i,S}$.
While the second equation characterizes the relationship of $B_{i,S}$ between any pair of service providers, as a result the above equation system is essentially a linear equation system with $N$ unknowns and $N$ independent linear equations. Thus if there is
a solution to the equation system, it must be unique.

%%%Appendix G
\section{MSNE Conditions and Properties} \label{Appen:MSNE Conditions}
The bandwidth allocation at an MSNE can be computed via the following system of equations:
\begin{displaymath}
\left\{
\begin{array}{ll}
\lambda_S[   Nu^{'}(R_S)+\frac{\lambda_S\sum\limits_{i\in \mathcal{N}} B_{i,S}R_0}{N_f} u^{''}(R_S)    ]\\
 = Nu^{'}(R_M)+R_Mu^{''}(R_M)\\
\frac{\lambda_S^2\Delta B_{ij,S}}{N_f}u^{''}(R_S)=\frac{\Delta B_{ij,M}}{N_m}u^{''}(R_M)
\end{array}
\right.
\end{displaymath}
where $\Delta B_{ij,S}=B_{i,S}-B_{j,S}$ is the difference of bandwidth allocation to small-cells between SP $i$ and SP $j$, the same
for $\Delta B_{ij,M}$.

Proposition~\ref{Prop:Condition for MNE} results in the following corollary in the symmetric setting where all SPs have the same bandwidths, i.e., $B_i\equiv B$ for all $i\in \mathcal{N}$.
\begin{corollary}\label{Cor:SymmetricCase}
If all SPs have the same bandwidths $B$ and
\begin{align}
C_U \geq R_0 NB\left(\frac{\lambda_S}{1-\frac{\alpha}{N}}\right)^{\frac{1}{\alpha}} \frac{ \kappa N_f }{N_m},
\end{align}
then we have an MNE with
\begin{align}
\begin{split}
&K_M=N_m, K_S=0, K_U=N_f, \\
&R_M=\frac{N B R_0}{N_m}, R_S=0, R_U=\frac{C_U}{N_f}.
\end{split}
\end{align}
Otherwise we have an MSNE in which all SPs have the same bandwidth allocation
and can be determined by solving the following equations
\begin{align}
\begin{split}
& \bigg(1-\frac{\alpha}{N}\bigg) \left(\frac{N_m}{N B_M R_0} \right)^{\alpha}  =  \left(\frac{\kappa N_f}{\kappa N B_S \lambda_S R_0 + C_U} \right)^\alpha \\
& \qquad \qquad \qquad \qquad \times \lambda_S \left(1-\alpha \frac{\kappa \lambda_S B_S R_0}{\kappa N B_S \lambda_S R_0 + C_U} \right) \\
& B_S+B_M   = B, \qquad B_S, B_M > 0.
\end{split}
\end{align}
\end{corollary}

Using Corollary~\ref{Cor:SymmetricCase} we can analyze the asymptotic scenario of $N\rightarrow\infty$ as given below.
\begin{proposition}\label{Prop:fullprop4}
Assume that the total amount of bandwidth of $N$ SPs is fixed to be $B$ and each SP $i$ gets the same proportion $B_i=\frac{B}{N}$. If we have
\begin{align}
B_U \lambda_U > B \frac{\kappa N_f \lambda_S^{\frac{1}{\alpha}}}{N_m},
\end{align}
then there exists an $N^*(B_U \lambda_U)$ such that for all $N > N^*(B_U \lambda_U)$ we have an MNE so that $B_{i,M}=B/N$ and $B_{i,S}=0$ for all $i\in \mathcal{N}$.
Otherwise we always have an MSNE, with $\lim_{N\rightarrow \infty} (N B_{i,M}, N B_{i,S})=(B_M, B_S)$ for all $i$ with
\begin{align}
B_S =  \frac{B- \frac{B_U \lambda_U N_m}{\kappa N_f \lambda_S^{\frac{1}{\alpha}} }}{1+\frac{\lambda_S N_m}{\lambda_S^{\frac{1}{\alpha}} N_f}}, B_M =  \frac{\bigg(B+B_U \frac{\lambda_U}{\kappa \lambda_S} \bigg) \frac{\lambda_S N_m}{\lambda_S^{\frac{1}{\alpha}} N_f}}{1+\frac{\lambda_S N_m}{\lambda_S^{\frac{1}{\alpha}} N_f}}.
\end{align}
Furthermore,
\begin{align}
\begin{split}
& K_S=\frac{\kappa B_S \lambda_S N_f}{\kappa B_S \lambda_S+B_U\lambda_U}, K_U=\frac{B_U\lambda_U N_f}{\kappa B_S \lambda_S+B_U\lambda_U}, \\ &
K_M=N_m, R_S = \frac{R_0(\kappa B_S \lambda_S+B_U\lambda_U)}{\kappa N_f},\\
&  R_M=\frac{B_M R_0}{N_m}, R_U = \frac{R_0(\kappa B_S \lambda_S+B_U\lambda_U)}{N_f}.
\end{split}
\end{align}
\end{proposition}

%If all SPs have the same bandwidths $B$, then an MNE implies
%\begin{align}
%\begin{split}
%&K_M=N_m, K_S=0, K_U=N_f, \\
%&R_M=\frac{N B R_0}{N_m}, R_S=0, R_U=\frac{C_U}{N_f}.
%\end{split}
%\end{align}
%Otherwise we have an MSNE in which all SPs have the same bandwidth allocation
%and can be determined by solving the system of equations above.

%\begin{displaymath}
%\left\{
%\begin{array}{ll}
%&\lambda_S\Big[  N \big( \frac{\kappa N_f}{\kappa N\lambda_S B_{i,s}R_0+C_U}\big )^{\alpha} - \alpha \frac{N\lambda_SB_{i,s}R_0}{N_f}\\
%&\big( \frac{\kappa N_f}{\kappa N\lambda_S B_{i,s}R_0+C_U}\big )^{\alpha+1}   \Big]
%=(N-\alpha)\big( \frac{N_m}{NB_{i,M}R_0}  \big)^{\alpha}\\
%&B_{S}+B_{M}=B
%\end{array}
%\right.
%\end{displaymath}

%%%Appendix H
\section{Proof of Proposition \ref{Prop:MSNE Bandwidth Comparison}} \label{Appen: MSNE Bandwidth Comparison}
The equilibrium equations for MSNE in scenarios with and without unlicensed spectrum are given below:
\begin{align}
\notag &\text{With}:
\lambda_S\Big[  Nu^{'}(R_S)+\frac{\lambda_SB_SR_0}{N_f}u^{''}(R_S)  \Big] =\\
&Nu^{'}(R_M)+R_Mu^{''}(R_M)\\
\notag &\text{Without}:
\lambda_S\Big[  Nu^{'}(R_S)+R_Su^{''}(R_S)  \Big] = \\
&Nu^{'}(R_M)+R_Mu^{''}(R_M) \label{Eqn:MSNE equations without unlincesed}
\end{align}

We only need to compare the LHS for two scenarios:
\begin{align}
A_1=&N\big( \frac{\kappa N_f}{C_U+\kappa\lambda_SB_SR_0} \big)^{\alpha}-\alpha\frac{\lambda_SB_SR_0}{N_f}\\
&\big(  \frac{\kappa N_f}{C_U+\kappa\lambda_SB_SR_0}  \big)^{\alpha+1}\\
A_2=&N\big( \frac{N_f}{\lambda_SB_SR_0} \big)^{\alpha}-\alpha\big( \frac{N_f}{\lambda_SB_SR_0} \big)^{\alpha}
\end{align}

Doing the calculation of $A_1-A_2$, the resulting ratio's numerator $Z$ can be simplified to:
\begin{align}
\notag Z=&(\kappa\lambda_SB_SR_0)^{\alpha}(C_U+\kappa\lambda_SB_SR_0)-\alpha(\kappa\lambda_SB_SR_0)^{\alpha+1}\\
&-(N-\alpha)(C_U+\kappa\lambda_SB_SR_0)^{\alpha+1}
\end{align}

Taking the derivative with respect to $C_U$, we have:
\begin{equation}
\frac{\partial Z}{\partial C_U}=(\kappa\lambda_SB_SR_0)^{\alpha}-(N-\alpha)(1+\alpha)(C_U+\kappa\lambda_SB_SR_0)^{\alpha}
\end{equation}

Evaluate the expression at $C_U=0$:
\begin{align}
\notag \frac{\partial Z}{\partial C_U}|_{C_U=0}=&(\kappa\lambda_SB_SR_0)^{\alpha}(1-(N-\alpha)(1+\alpha))\\
\notag &\le(\kappa\lambda_SB_SR_0)^{\alpha}(1-(2-\alpha)(1+\alpha))\\
&=(\kappa\lambda_SB_SR_0)^{\alpha}\big[ (\alpha-\frac{1}{2})^2-\frac{5}{4} \big] <0
\end{align}

Therefore we can conclude that $\frac{\partial Z}{\partial C_U}<0, C_U\ge 0$. On the other hand, we have $Z=0$ when
$C_U=0$. As a result, $A_1<A_2$ always holds.

Thus, we can conclude at MSNE, the optimal bandwidth allocation to small-cells with unlicensed spectrum is always less than
that of without unlicensed spectrum.

\section{Proof of Results in Section \ref{Sec:Licensed vs Unlicensed Spectrum}} \label{Appen:Licensed vs Unlicensed Spectrum}
For ideal case, it is a simple allocation optimization and we only need to equalize the marginal utility increase with respect to bandwidth increase for both mobile users and fixed users. Since both WiFi network in unlicensed spectrum and small-cells in licensed spectrum are able to serve fixed users, the social planner should always choose the one which can lead to more rate to invest bandwidth. As a result, when $\lambda_S>\lambda_U$, $B_U^{\text{opt}}=0$, when $\lambda_S<\lambda_U$, $B_S^{\text{opt}}=0$, when $\lambda_S=\lambda_U$, what matters is only the total bandwidth allocated to small-cells and unlicensed spectrum, while the split between them is arbitrary.

When $\lambda_S>\lambda_U$, the optimal bandwidth allocation strategy is given by:
\begin{displaymath}
\left\{
\begin{array}{ll}
\lambda_Su^{'}(\frac{\lambda_SB_S^{\text{opt}}R_0}{N_f})=u^{'}(\frac{B_M^{\text{opt}}R_0}{N_m})\\
B_S^{\text{opt}}+B_M^{\text{opt}}=B
\end{array}
\right.
\end{displaymath}

When $\lambda_S<\lambda_U$, the optimal bandwidth allocation strategy is given by:
\begin{displaymath}
\left\{
\begin{array}{ll}
\lambda_Uu^{'}(\frac{\lambda_UB_U^{\text{opt}}R_0}{N_f})=u^{'}(\frac{B_M^{\text{opt}}R_0}{N_m})\\
B_U^{\text{opt}}+B_M^{\text{opt}}=B
\end{array}
\right.
\end{displaymath}

When $\lambda_S=\lambda_U=\lambda$, the optimal bandwidth allocation strategy is given by:
\begin{displaymath}
\left\{
\begin{array}{ll}
\lambda u^{'}(\frac{\lambda(B_S^{\text{opt}}+B_U^{\text{opt}})R_0}{N_f})=u^{'}(\frac{B_M^{\text{opt}}R_0}{N_m})\\
B_S^{\text{opt}}+B_U^{\text{opt}}+B_M^{\text{opt}}=B
\end{array}
\right.
\end{displaymath}

For practical scenario, when $\lambda_S\ge \lambda_U$, it's always optimal for the social planner to allocate all bandwidth to licensed spectrum because for $\alpha$-fair utility functions, maximizing revenue is exactly the same as maximizing social welfare for monopoly scenario. In competitive scenario, it's also easy to verify for $\alpha$-fair utility functions, equation (\ref{Eqn:MSNE equations without unlincesed}) would also yield the same bandwidth allocation which maximizes social welfare.

When $\lambda_S=\lambda_U=\lambda$, another optimal allocation occurs at the point that the social planner allocate $B_U=B_S^{\text{opt}}+B_U^{\text{opt}}$ to unlicensed spectrum and $B_L=B_M^{\text{opt}}$ to licensed spectrum. We can prove that in this case the SP(s) would only allocate the bandwidth $B_L$ to macro-cells under certain conditions, which therefore achieves the same social welfare as the benchmark optimal case. To prove this, we notice that the condition for the SP(s) to only allocate bandwidth to macro-cells is given by equation (\ref{Eqn:Condition for MNE}):
\begin{equation}\label{Eqn:Condition for MNE appen}
\lambda_S\big(  \frac{C_U}{\kappa N_f} \big)^{-\alpha} \le \big( \frac{\sum\limits_{i=1}^{N}B_iR_0 }{N_m}\big)^{-\alpha} -\alpha \frac{B_{i_{\max}}R_0}{N_m} \big( \frac{\sum\limits_{i=1}^{N}B_iR_0 }{N_m}\big)^{-\alpha-1}
\end{equation}

When $B_U=B_S^{\text{opt}}+B_U^{\text{opt}}$, $B_L=B_M^{\text{opt}}$, we have:
\begin{equation}
\frac{C_U}{N_f}=\lambda^{\frac{1}{\alpha}}\frac{B_M^{\text{opt}}R_0}{N_m}
\end{equation}

Therefore we have:
\begin{align}
&\lambda_S\big(  \frac{C_U}{\kappa N_f} \big)^{-\alpha}=\kappa^{\alpha}\big(  \frac{B_M^{\text{opt}}R_0}{N_m}  \big)^{-\alpha}\\
\notag &\big( \frac{B_LR_0 }{N_m}\big)^{-\alpha} -\alpha \frac{B_{i_{\max}}R_0}{N_m} \big( \frac{B_LR_0 }{N_m}\big)^{-\alpha-1}\\
&=(1-\frac{\alpha}{N})\big( \frac{B_LR_0 }{N_m}\big)^{-\alpha}
\end{align}
which means (\ref{Eqn:Condition for MNE appen}) holds when it satisfies the following:
\begin{equation}
\kappa^{\alpha}+\frac{\alpha}{N}\le 1
\end{equation}

It's easy to verify when $N\ge 2$, the above condition is always satisfied for $\alpha\in (0,1)$. When $N=1$, it is satisfied when
$\alpha\in (0,0.5]$ .

We can also prove that when $\lambda_S<\lambda_U$, one possible way to achieve the optimal benchmark social welfare is to allocate $B_U=B_U^{\text{opt}}$ to unlicensed spectrum and $B_L=B_M^{\text{opt}}$ to licensed spectrum. As a result, under certain conditions, the SP(s) would also allocate all bandwidth $B_L$ only to macro-cells, which therefore leads to the same social welfare as in scenario 1). By similar argument, the conditions for this to hold is the following:
\begin{equation}
\kappa^{\alpha}\frac{\lambda_S}{\lambda_U}+\frac{\alpha}{N}\le 1
\end{equation}

It's easy to verify when $N\ge 2$, the above condition is always satisfied for $\alpha\in (0,1)$. When $N=1$, it is satisfied when
$\alpha\in (0,\alpha_0]$, where $\alpha_0$ is the unique solution to the following equation: $$\kappa^{\alpha_0}\frac{\lambda_S}{\lambda_U}+\alpha_0 = 1.$$

\end{appendices}

\end{document}